\documentclass[preprint]{aastex}

\begin{document}

\title{Faint High-Latitude Carbon Stars Discovered by the Sloan Digital Sky
Survey: An Initial Catalog}

\author{Ronald A. Downes\altaffilmark{1}, 
Bruce Margon\altaffilmark{1}, 
Scott F. Anderson\altaffilmark{2},
Hugh C. Harris\altaffilmark{3},
G. R. Knapp\altaffilmark{4},
Josh Schroeder\altaffilmark{4},
Donald P. Schneider\altaffilmark{5},
Donald. G. York\altaffilmark{6},
Jeffery R. Pier\altaffilmark{3}, and
J. Brinkman\altaffilmark{7}
}

\altaffiltext{1}{Space Telescope Science Institute, 3700 San Martin Drive, Baltimore, MD 21218}
\altaffiltext{2}{Department of Astronomy, University of Washington, Box 351580,
Seattle, WA 98195-1580}
\altaffiltext{3}{US Naval Observatory, Flagstaff Station, P.O. Box 1149, Flagstaff, AZ
86002-1149}
\altaffiltext{4}{Princeton University Observatory, Peyton Hall, Princeton, NJ 08544-1001}
\altaffiltext{5}{Department of Astronomy and Physics, 525 Davey Laboratory, Pennsylvania State University, University Park, PA 16802}
\altaffiltext{6}{University of Chicago, Astronomy \& Astrophysics Center, 5640 South Ellis Avenue, Chicago, IL 60637}
\altaffiltext{7}{Apache Point Observatory, P.O. Box 59, Sunspot, NM 88349-0059}

\begin{abstract}

A search of more than 3,000 square degrees of high latitude sky by the
Sloan Digital Sky Survey has yielded 251 faint high-latitude carbon
stars (FHLCs), the large majority previously uncataloged. We present
homogeneous spectroscopy, photometry, and astrometry for the
sample. The objects lie in the $15.6 < {\it r} < 20.8$ range, and
exhibit a wide variety of apparent photospheric temperatures, ranging
from spectral types near M to as early as F. Proper motion
measurements for 222 of the objects show that at least $50\%$, and
quite probably more than $60\%$, of these objects are actually low
luminosity dwarf carbon (dC) stars, in agreement with a variety of
recent, more limited investigations which show that such objects are
the numerically dominant type of star with C$_{2}$ in the
spectrum. This SDSS homogeneous sample of $\sim110$ dC stars now
constitutes $90\%$ of all known carbon dwarfs, and will grow by
another factor of 2-3 by the completion of the Survey. As the spectra
of the dC and the faint halo giant C stars are very similar (at least
at spectral resolution of $10^3$) despite a difference of 10 mag in
luminosity, it is imperative that simple luminosity discriminants
other than proper motion be developed. We use our enlarged sample of
FHLCs to examine a variety of possible luminosity criteria, including
many previously suggested, and find that, with certain important
caveats, {\it JHK} photometry may segregate dwarfs and giants.

\end{abstract}


\section{Introduction}

Carbon stars (objects with prominent C$_{2}$ in their spectra) have
been studied for more than a century \citep{dun84}, although faint (R $> 13$)
high-latitude carbon stars (FHLCs), which presently number in the
hundreds, were not easily found until recently.  Most FHLCs are
thought to be distant giants, as there is no obvious way for C$_{2}$
to reach the photosphere prior to the red giant phase. While
relatively rare, these objects are interesting because, for example,
of their utility as a halo velocity tracers \citep{msg85,bem91}.  
However, a small number of FHLCs display parallaxes
and/or large proper motions, implying main-sequence luminosities
(M$_{V} \sim10$), and have been designated dwarf Carbon (dC) stars.
It now appears that a significant fraction of FHLCs are in fact not
distant giants, but are nearby dwarfs \citep[hereafter Paper I]{mar02}.  Based
on data obtained during the Sloan Digital Sky Survey \citep[SDSS;][]{york00}
commissioning period, we claimed in Paper I that at
least $40-50\%$ of all FHLCs are dwarfs. 

Here we report the results of an expanded sample of SDSS FHLCs.  In
Section~2, we discuss the observations and selection criteria for our
sample, while Section~3 describes the classification of the objects as
dwarfs or giants.  A key problem remains the derivation of simple
luminosity discriminants other than proper motion, as the low
resolution ($\sim1000$) spectra of the giants and dwarfs are so
similar.  Section~4 discusses photometric and spectroscopic luminosity
indicators, while our conclusions are given in Section~5.

\section{Observations}

The FHLC candidates were chosen by analyzing the imaging data from the
SDSS camera \citep{gunn98,lgi01}, which obtains images in
%five bands ({\it u, g, r, i, z}; \citet{fig96}) almost simultaneously.
five bands \citep[{\it u, g, r, i, z};][]{fig96} almost simultaneously.
The instrumental fluxes are calibrated via a network of primary and
secondary standard stars \citep{fig96,hog01,smi02}.  
Observations of previously known FHLCs in the SDSS
system \citep{kms98} were used to determine approximately where in the
SDSS color-color diagrams FHLCs were expected, and an automated analysis of
the photometric database was used to identify objects that appear in
those regions. More details on object selection can be found in
Paper~I.

Spectroscopic observations were then obtained for as many of the FHLC
candidates as possible.  The spectra, obtained with a SDSS fiber-fed CCD
spectrograph, cover the wavelength range 3800~\AA-9200~\AA~with
$\Delta\lambda$/$\lambda \sim1800$; see \citet{sto02} for
details. Although more than 600 spectra are obtained in each
observation, our survey is not 100$\%$ complete due to two factors.
First, there is a minimum separation of the fibers, so a FHLC
candidate close to a higher priority science target cannot be
observed.  Second, there are regions where all fibers are utilized by
unrelated higher priority programs.  Nonetheless, the SDSS FHLC survey
does provide a large homogeneous sample of objects.

As was the case in Paper~I, we have not relied on an automated
algorithm to select true FHLC stars from the SDSS spectra, but rather
visually examined all spectra. The SDSS First Data Release DR1
\citep{aba03} contains spectra of about 500 unresolved sources that the
SDSS photometric selection flagged as a possible FHLC.
However, manual examination of these spectra shows that only
$\sim10$\% actually display C$_2$ bands.  The bulk of the
contaminants are late type stars, and a few are high redshift quasars, where
prominent L$\alpha$ emission simulates a band head. In this respect
the contaminants are identical to those in C star surveys of decades past,
made on objective prism plates.  Clearly, the efficiency
of photometric selection of FHLCs is modest, due to color degeneracy
with both interesting and uninteresting objects, Galactic and
extragalactic.  However, the SDSS will at completion still derive a
very large, homogeneous sample of FHLCs, in part because multiple
other unrelated, high-priority scientific projects in the survey, in
particular the search for distant QSOs, target many of these
candidates for spectroscopy due to the color degeneracy.  Even though
such programs may have relatively small contamination rates by FHLCs,
they target very large numbers of objects.

On the basis of the spectroscopic observations, 251 of the SDSS
candidates (including the 39 listed in Paper~I, which are repeated
here for convenience and completeness) were found to be FHLCs.  Ninety
seven have data publically available in the DR1, and the remainder are
newly presented here.  The spectra of all FHLCs can be found via
anonymous ftp\footnote{\bf ftp.stsci.edu; cd pub/science/SDSS\_Carbon}; sample
spectra are displayed in Paper~I.  The 251 objects are listed in
Table~\ref{tbl-1}, which provides astrometric (coordinates and, when
available, proper motions) and photometric (both Sloan and, when
available, 2MASS colors) data.  We also note a luminosity class (giant
or dwarf) based on the SDSS-derived proper motions (see Paper~I for
details): all objects with $\>3\sigma$ detections of motion are
considered dwarfs.  The spectra are qualitatively similar to those
noted in Paper~I, and we continue to see the broad range of NaD
strength (presumptively temperature) previously noted. However, the
fraction of objects with noticeable Balmer emission (6$\%$) is only
about half that seen in our initial sample.

A few special cases of extreme color are worthy of mention.
In Paper~I, no N-type stars were detected by the selection software
(the 2 objects shown there were found in an SDSS vs. 2MASS comparison),
although we noted that their extreme red colors should make them easier to
find than the R-type.  Our expanded survey has resulted in the
detection of one (but only one!) N-type star - SDSS J125149.87+013001.8.
A small fraction of the sample are exceptionally blue, and show Balmer and
Ca~II absorption in addition to the C$_{2}$ bands.  These ``F/G Carbon
Stars'', discussed elsewhere \citep{sch02,knapp03}, were initially found
serendipitously.  Searches of the SDSS standard star sample, selected
by colors optimized for metal-poor F-subdwarfs, then added $\sim20$
more cases.  This small fraction of very red or very blue objects, not
found by our nominal FHLC target selection software, stands as yet 
another warning of incompleteness in this survey.

Figures 1-3 display color-color diagrams of the SDSS FHLCs, and define
the location in the diagrams where FHLCs are found. The large scatter
in {\it (u-g)} is due to the lower sensitivity of the survey in the {\it
u} band combined with the faintness of the objects; however, the one
N-type star is clearly separated from the other objects.  There is a
tight correlation in the {\it (g-r),(r-i)} diagram, with the one
N-type star again distinctly visible.  Also, the ``F/G Carbon Stars''
are clearly distinct, but appear to be an extension of
the R-type star correlation.  Similarly, in the {\it (r-i),(i-z)}
diagram, the stars are fairly closely clumped, with the ``F/G Carbon Stars''
again following the R-type stars; the N-type star is not as clearly
separated from the R-types in this diagram.

In Paper~I, we noted that our SDSS survey is complementary to the
2MASS survey, as the former will detect the warmer R-type stars, while
the latter will be useful for detecting the cooler N-type stars
\footnote{It is of course unclear whether 2MASS alone, which
lacks spectroscopic data, can be used to select FHLCs with any
reasonable efficiency.}.  It was also noted that the SDSS survey
detects objects too faint for 2MASS.  We have searched the 2MASS
All-Sky Data Release Point Source Catalog for each of the SDSS FHLCs,
and find about $50\%$ of the objects are detected; the 2MASS colors
are listed in Table~\ref{tbl-1}.  Figure~\ref{fig4} shows the
distribution of 2MASS detections in SDSS {\it r} magnitude bins.  Not
surprisingly, most of the bright objects ({\it r} $< 18.0$) are
detected in 2MASS, while most of the faint objects ({\it r} $> 19.0$)
are not.

To support our search for luminosity discriminators, we measured
equivalent widths for both the discrete lines (the Balmer lines, NaD),
and the strong CaH bands ($\lambda6382$ and $\lambda6750$) in each
spectrum in which they were present.  For the C$_{2}$ ($\lambda4737$,
$\lambda5165$, $\lambda5636$, and $\lambda6191$) and CN
($\lambda7900$) band heads, we measured a band head intensity
corresponding to the average flux at the peak of the band divided by
the average flux at the bottom of the band.

\section{Analysis}
\subsection{Luminosity Determination}

The SDSS Astrometric Pipeline \citep{pmh02} computes J2000.0 positions
for all detected objects, and these positions have been compared with
USNO-B data to obtain proper motions for almost all the FHLCs (see
Table~\ref{tbl-1}); those without motions are either too faint to
obtain reliable positions, or are in crowded fields.  All objects with
non-zero motions significant at a $3\sigma$ level are considered dwarfs;
see Paper~I for a discussion of the validity of using only proper
motion to determine the luminosity class.

Those objects that do not show motion could either be giants or
distant dwarfs whose motions are too small to detect.  A quick
examination at the magnitudes of the objects without detected motion
shows that only a handful are fainter than the dwarfs, so these
objects are likely a combination of true giants and distant dwarfs.

To determine which objects are close enough (if dwarfs) for a motion
to be detectable, we examined the distribution of observed motions,
and find that most (over $90\%$) have proper motions $\ge 20$~mas
yr$^{-1}$, which we take as our minimum detectable value.  After
converting the SDSS magnitudes to V \citep{fig96}, and assuming
M$_{V}=+10$, we converted all proper motions to a linear scale, and
find that $\sim90\%$ of the objects have tangential velocities
$\ge50$~km s$^{-1}$.  Finally, we determined the V magnitude an object
would have if it were at these limits (i.e., would be barely
detectable), and find that V$_{min}$~=~18.6.  This means that any
object brighter than this limit is likely a true giant, which is
$32\%$ (35) of the objects without observed motions; the nature of the
remaining objects must be considered uncertain.  The occasional
renegade dwarf may of course by chance have a small tangential
component to its motion.  We exclude from this analysis the ``F/G
Carbon Stars'', which are likely more luminous than the fiducial
M$_{V}=+10$, and label them all of uncertain luminosity in
Table~\ref{tbl-1}, unless a positive detection of proper motion is
present.

For the remaining 74 objects (excluding the 29 objects that had no
proper motion data available), it is possible to make a statistical
argument concerning the number that should be giants.  The faintest
star in our sample has V=21.3, and it would need to have a tangential
velocity of 175~km s$^{-1}$ to be detectable.  About one-third of the
dwarfs have a motion above this limit, so $0.33 \times 74~=~24$ of the
uncertain objects should be dwarfs with motions large enough to
detect, and hence the lack of detected motion implies they are true
giants (although we don't know which ones).  Thus, about half of the
objects without detected motions are probably giants.

A different conclusion emerges, however, if a significant fraction of
the sample is made of disk dwarfs with space velocities smaller than
velocities of halo stars.  They will sometimes have tangential
velocities less than 50 km s$^{-1}$, so some will contaminate the brighter
stars that were called giants above.  They will all have tangential
velocites less than 175 km s$^{-1}$, so they may dominate the fainter stars
without significant motions, leaving fewer of the fainter uncertain
objects as implied giants than the 24 estimated above.  In this case,
significantly fewer than half of the objects without detected motions
would be giants.  The population models made for Paper~I indicated
that disk dwarfs were required to explain the stars with small radial
velocities.  A further analysis of the population mix will be deferred
to a future paper.

The total number of dwarfs in our survey (excluding the ``F/G Carbon stars'')
is 107, while the total number of non-dwarfs (excluding the ``F/G Carbon
stars'' and the 3 objects in the Draco dwarf galaxy) is 117.  Thus, at
least $48\%$ of the FHLCs are dwarfs, consistent with the estimate of
$\ge50\%$ of Paper~I.  If we assume half the uncertain objects are
dwarfs, then at least $63\%$ of the FHLCs are dwarfs.  The recent summary of
known dC stars by \citet{lkr03} lists 14 objects in total not due to
SDSS.  Thus, the homogeneous sample of SDSS stars in Table~\ref{tbl-1}
that are unambiguously dwarfs now constitutes $90\%$ of all known dC stars.

In Paper~I, we derived a surface density of $\>0.05$~deg$^{2}$, which
was consistent with the value of $0.072 \pm 0.005$~deg$^{2}$ of
\citet{cgw01}, derived from a bright, photographic sample.  With the
same caveats about the completeness of our sample (see Paper~I), we
derived a surface density from our expanded sample (224 objects in
$\sim$3600~deg${^2}$) of 0.06~deg$^{2}$ (again excluding the ``F/G
Carbon stars'' and the 3 Draco stars).  It is clear that careful
incompleteness corrections, not currently available, will be needed to
understand the true surface density of the faintest FHLCs.

\section{Discussion}

As previously noted, at least $50\%$ the FHLCs are 
dwarfs.  Therefore the development of a simple
observational luminosity discriminant is imperative if we are to, for
example, use the giants as halo velocity tracers.  The fact that the
dwarfs and giants differ in intrinsic luminosity by $\sim10$ mag gives
hope that there would be such discriminants.  In fact, several
luminosity discriminants have been proposed, although the small number
of objects known to date has made it difficult to assess the validity
and range of applicability of these discriminants.  With our large
sample of objects, an objective analysis of the proposed
discriminants, as well as the possibility of uncovering others, is now
possible.

\subsection{Previously proposed photometric luminosity discriminants}

Numerous authors \citep[e.g.,][]{gma92,joyce98,tiw00,lkr03} 
have proposed near-IR colors as a luminosity
discriminant, while \citet{mac03} showed a tight sequence for N-type
stars for his survey objects in the {\it (H-K)}, {\it (J-H)} diagram.
While we will discuss the luminosity aspect below, we present in
Figure~\ref{fig5} the 2MASS color-color diagram for all stars
($\sim100$ stars) in our survey, in the carbon star catalog ($\vert b
\vert > 10^\circ$; $\sim300$ stars) of \citet{alk01}, and all previously
reported FHLC dwarfs (11 stars), summarized by \citet{lkr03}.

While the N-type stars continue to show the trend seen by \citet{mac03},
the R-type stars are far more scattered, and the ``F/G Carbon stars''
appear to form a nice continuation of the C stars sequence.  The one
N-type star in our survey fits nicely in with those from \citet{alk01}, 
while the one R-type star from \citet{alk01} at
{\it (H-K)}=1.0 is classified as R-N, so is likely an N-type; the R-type
star at {\it (H-K)}=1.6 is the bright Mira variable RU~Vir.

Although near-IR colors have been proposed as a luminosity
discriminant, \citet{wj96} and \citet{jbh98} argue on purely physical
grounds that {\it JHK} colors alone should not be an unambiguous indicator.
Based on 4 objects from Paper~I, we found that 2 objects were
consistent with the {\it JHK} discriminant, while 2 objects were not
consistent.  In Figure~\ref{fig6}, we show an enlargement of the
near-IR color-color diagram, with just the SDSS stars, as well as the
previously known dwarfs, displayed.  For {\it (H-K)} colors bluer than about
0.3, there is no separation between giants and dwarfs.  However, for
redder colors, the bulk of the objects are dwarfs.  For the (somewhat
arbitrary) line shown in the figure, 20 ($74\%$) of the objects are
dwarfs, 1 ($4\%$) is a giant, and 6 ($22\%$) are uncertain; if only
the confirmed giants and dwarfs are counted, then $95\%$ of the
objects are dwarfs.  A slight shift of the line to the blue (or red)
does not significantly change the giant to dwarf ratio. An examination
of Figure~\ref{fig5} shows that most of the previously known dwarfs do
indeed appear to be offset from the bulk of the R-type stars from the
\citet{alk01} catalog, but with the addition of the SDSS FHLCs, it now
appears that this discriminant can be refined and confirmed.

\citet{lkr03} proposed that dwarfs and giants could be separated in
the {\it (J-K)}, {\it (R-J)} diagram.  While we do not have R
magnitudes for our SDSS sample, the SDSS {\it r} band should be a
reasonable proxy to the Johnson R, so we have examined this
discriminant with our sample.  Figure~\ref{fig7} shows that, as with
the near-IR colors, this discriminant can identify dwarfs, but is
somewhat less effective (there are 5 giants in the dwarf region) than
using only 2MASS colors.

In Paper~I, we commented that photometric variability may prove to be
a simple luminosity criterion, since only giants should show
consistent, chaotic variations associated with mass loss; absence of
variability, however, does not imply the object is a dwarf.  Many of
the SDSS objects have multiple photometric measurements in the imaging
database, so we have performed a preliminary check for variability.
Given the faintness of many of the objects (particularly in {\it u}),
we define an object as variable if at least 2 of the {\it g, r}, and
{\it i} magnitudes changed by at least 0.10 mag.  With that criteria,
only 1 of the 49 FHLCs with multiple measurements is variable, and
this object is listed in Table~\ref{tbl-2}.  That object has an
uncertain luminosity class, and thus is a strong candidate for a
giant.  Note, however, that the object could be similar to the dC
PG~0824+289 \citep{hbj93}, which shows variability due to heating
effects from a hot companion, and so multiple measurements to look for
periodic variability would be interesting.

\subsection{New photometric luminosity discriminants}

In Figures 1-3, we plot the 3 SDSS
color-color diagrams. An examination of those figures shows that,
while the ``F/G Carbon stars'' are clearly separated from the R-type stars,
there is no indication of a luminosity discriminant. We also
investigated other color-color plots, none of which presented any
valid discriminant.

The model for a dwarf C star is that the C$_{2}$ was deposited in a
previous mass-transfer episode from a companion that is now a faint
white dwarf.  In some cases (e.g., PG~0824+289 \citep{hbj93}; SBS
1517+5017 \citep{lsl94}), the white dwarf is hot enough to be
detectable in the visible.  However, in most cases, the white dwarf is
too cool to be visible in the optical.  We examined those dwarfs in
our sample that have the bluest {\it u-g} colors, to look for any
spectroscopic evidence of a white dwarf, but none was found. On the
other hand, it may be possible to detect the white dwarf in the
ultraviolet, where the C star is faint and the contrast may be
optimized.

To determine how faint (i.e., cool) the white dwarf would typically
have to be to be undetected either spectroscopically or
photometrically, we selected one blue and one red giant FHLC from our
sample, and added a blackbody and/or real white dwarf spectrum of
various temperatures.  We find that a 10,000~K blackbody is barely
detectable in the red giant, while an 11,000~K blackbody is barely
detectable in the blue giant.  However, in the satellite ultraviolet,
these cool white dwarfs could be detectable (e.g., with the ACS SBC
camera on {\it HST}).  Therefore, UV colors may in fact be a useful
photometric luminosity discriminant.

\subsection{Previously proposed spectroscopic luminosity discriminants}

Numerous spectroscopic luminosity discriminants have been previously
proposed.  \citet{gma92} proposed that the appearance of a strong
C$_{2}$ band head at $\lambda6191$ indicated the object was a dwarf,
while \citet{mar02} suggested the feature was both temperature and
luminosity dependent.  However, both studies noted that while the
absence of $\lambda6191$ could not be used to firmly classify the
object as a giant, the presence of $\lambda6191$ was a good indicator
that the object is a dwarf.  In Figure~\ref{fig8} we show a histogram
of the strength of $\lambda6191$, which demonstrates that this feature is in
fact probably not a good luminosity discriminant.  When $\lambda6191$
is weak (1.0-1.4), about $50\%$ of the objects are confirmed dwarfs,
while at its strongest ($>1.6$) about a third of the objects are
confirmed dwarfs.  Only in the 1.4-1.6 bin do the dwarfs dominate
($70\%$).  In total, $50\%$ of the objects with $\lambda6191$ are
confirmed dwarfs, so this feature can no longer be considered a
luminosity discriminant.

In Paper~I we suggested that the presence of H$\alpha$ emission could
be an indicator that the object is a giant, since dwarfs are unlikely
to process active chromospheres or undergo active mass loss.  However,
a contrary case (PG~0824+289) was noted, where heating of the dwarf by
a hot DA companion caused the emission.  In that case, the white dwarf
was visible in the optical spectrum, so H$\alpha$ emission in dCs
where there is no indication of a hot white dwarf could still be a
luminosity discriminant.  Of the 14 objects in our sample with
H$\alpha$ emission, 3 are giants, 3 are dwarfs, and the remaining 8
have are uncertain luminosity.  Thus, the use of H$\alpha$ emission as
a luminosity discriminant is speculative.  The dCs with H$\alpha$
emission are rare, and thus should be further observed.

The bands of CaH at $\lambda\lambda6382,~6750$ are normally strong
only in K and M dwarfs \citep[e.g.,][]{khm91}, which led us to suggest
in Paper~I that the presence of these features was a luminosity
indicator in FHLCs.  In Figure~\ref{fig9} we show plot of the
equivalent width of $\lambda6750$ versus that of NaD.  While it is
true that there are no giants whose spectra show CaH, there are 10
objects (out of 16 total) that have an uncertain luminosity.  So,
while promising, this discriminant must still be consider unproven.

It is interesting to note that sharp turn-on of the CaH feature at
W$_{\lambda}$(NaD)~=~10~\AA, although there are two objects (one with
weaker NaD (hotter) which shows CaH, one with stronger NaD (cooler)
that does not) that are exceptions.  To estimate the temperature of
this turn-on point, the temperature index \citep{coh79} of every star
that showed NaD was measured, and a fit of NaD as a function of Temperature
Index was made.  The Temperature index was
converted to a temperature \citep{yam67}, with the result that
the CaH turn-on occurs at T=2900~K.

\citet{gma92} suggested that dC's have strong $\lambda6191$ band heads
given their relatively weak CN band strength.  In Figure~\ref{fig10},
we plot the strength of $\lambda6191$ versus the strength of the CN
band at $\lambda7900$.  Unfortunately, there are so few confirmed
giants which show CN that a definitive statement about this criterion
is not possible.  However, given the range of $\lambda6191$ strengths
seen for the dwarfs alone, this discriminant can be considered doubtful.
On the other hand, the strength of the CN band alone, particularly at
the weakest and stronger levels, may be a discriminant (see below).

\subsection{New spectroscopic luminosity discriminants}

With our large sample of FHLCs, we have searched for possible new
spectroscopic discriminants.  \citet{cgw01} found a correlation
between the strengths of the C$_{2}$ bands at $\lambda5165$ and
$\lambda4737$, which they used for detection of carbon stars.  Our
data also shows this correlation, but there is no separation between
dwarfs, giants, and the stars of uncertain luminosity.  In a similar
vein, we show, in Figure~\ref{fig11}, the equivalent relation for the
strongest C$_{2}$ bands.  Again, there is a correlation, but C$_{2}$
band strength does not appear to be a luminosity discriminant.

As previously noted, the strength of the CN band at $\lambda7900$ may
be a luminosity indicator.  In Figure~\ref{fig12}, we plot a histogram
of the strength of this feature.  As can be seen, all objects with
band strengths greater than 1.6 are either giants (6) or of uncertain
luminosity (1; the object with the strongest CN).  Similarly, most of
the objects with weak CN are dwarfs (19); there are 2 giants and 1
object of uncertain luminosity.  Thus, extremely strong CN seems to
imply a giant luminosity, while extremely weak (but present) CN
implies a dwarf luminosity.

\section{Summary}

We have increased our SDSS survey of faint high-latitude carbon stars
to include 251 objects, at least $50\%$ of which are dwarfs.  Although
giant and dwarf carbon stars differ in luminosity by 10 mag, the
spectra (at the SDSS spectral resolution of $10^{3}$) are extremely
similar.  We have used our expanded sample to search for photometric
and spectroscopic discriminants, including an examination of those
previously proposed.

We find that SDSS colors and the strength of the C$_{^2}$ band head at
$\lambda6191$ are not good luminosity discriminants, while the presence of
H$\alpha$ emission as a discriminant is speculative. The
suggestion from Paper~I that the CaH bands at $\lambda\lambda6382,
6750$ indicate that the object is a dwarf must also still be considered
unproven.

The strength of the CN band at $\lambda7900$, when it is extremely
strong or extremely weak, appears to be a good luminosity
discriminant. However, it is of limited use, as almost half the
objects do not show this band.  For the redder objects, the use of
{\it JHK} colors appears to allow the identification of dwarfs.
Another possible discriminant is ultraviolet colors, where the
presumed white dwarf (which transferred its carbon to the now dC star)
may be detectable.

\acknowledgments

We thank J. MacConnell for helpful comments.  This publication makes
use of data products from the Two Micron All Sky Survey, which is a
joint project of the University of Massachusetts and the Infrared
Processing and Analysis Center/California Institute of Technology,
funded by NASA and the NSF. This research has also made use of the
SIMBAD database, operated at CDS, Strasbourg, France, and the
NASA/IPAC Infrared Science Archive, which is operated by the Jet
Propulsion Laboratory, California Institute of Technology, under
contract with the National Aeronautics and Space Administration.  

Funding for the creation and distribution of the SDSS Archive has been
provided by the Alfred P. Sloan Foundation, the Participating
Institutions, the National Aeronautics and Space Administration, the
National Science Foundation, the US Department of Energy, the Japanese
Monbukagakusho, and the Max Planck Society.  The SDSS Web site is
http://www.sdss.org/.

The SDSS is managed by the Astrophysical Research Consortium (ARC) for
the Participating Institutions.  The Participating Institutions are
The University of Chicago, Fermilab, the Institute for Advanced Study,
the Japan Participation Group, The Johns Hopkins University, Los
Alamos National Laboratory, the Max-Planck-Institute for Astronomy
(MPIA), the Max-Planck-Institute for Astrophysics (MPA), New Mexico
State University, University of Pittsburgh, Princeton University, the
US Naval Observatory, and the University of Washington.

\clearpage

\clearpage

\clearpage

\begin{deluxetable}{rcccccccrcrrl}
\tablenum{1}
\footnotesize
\rotate
\tablecaption{Faint High-Latitude Carbon Stars Discovered in SDSS\label{tbl-1}}
\tablewidth{0pt}
\tablehead{
\colhead{Name SDSS J+\tablenotemark{a}} & \colhead{Epoch\tablenotemark{b}} & \colhead{{\it r\tablenotemark{c}}} & \colhead{{\it u-g}} & \colhead{{\it g-r}}&
\colhead{{\it r-i}} &\colhead{{\it i-z}} &\colhead{J-H} &\colhead{H-K} & \colhead{r-J} &
\colhead{$\mu$\tablenotemark{d}} &\colhead{p.a.} &\colhead{Class\tablenotemark{e,f}} \\
& &  &  &  &  &  &  &  &  & \colhead{mas yr$^{-1}$} & \colhead{(deg)} 
}
\startdata

000354.23$-$104158.2 & 2000.74 & 18.59 & 2.78 & 1.36 & 0.45 & 0.31 & 0.69 & 0.03 & 1.92 & $43.4\pm 4.9$ & 110.1 & D \\ 
000643.14+155800.8 & 2001.72 & 19.80 & 2.11 & 1.68 & 0.52 & 0.12 &      &      &  & $41.8\pm6.6$ & 196.8 & D \\ 
001145.30$-$004710.2 &       & 18.15 & 3.16 & 1.47 & 0.57 & 0.24 & 0.58 & 0.56 & 2.06 & $4.4\pm4.7$ &  39.1 & U \\ 
001245.80$-$010521.9 & 1998.73 & 19.51 & 1.33 & 1.57 & 0.51 & 0.10 &      &      &  & $51.5\pm6.2$ & 175.8 & D \\ 
\smallskip
001716.46+143840.9 & 2001.72 & 18.36 & 2.49 & 1.58 & 0.51 & 0.25 & 0.57 & 0.53 & 1.95 &  $30.7\pm4.5$ & 149.9 & D \\ 
001836.23$-$110138.5 &       & 18.69 & 2.24 & 0.90 & 0.27 & 0.19 &      &      & &  $3.3\pm5.0$ &  10.9 & U \\ 
003013.09$-$003226.7&        & 19.44 & 3.81 & 1.89 & 0.47 & 0.28 &      &      & &  $8.7\pm6.5$ & 141.8 & U \\ 
003504.78+010845.9 &         & 17.28 & 4.17 & 1.91 & 0.68 & 0.50 & 0.89 & 0.45 & 2.66 &  $6.3\pm4.9$ & 331.5 & G$^{4}$ \\
003813.23+134551.1&          & 19.32 & 2.29 & 1.60 & 0.51 & 0.31 &      &      & &  $5.6\pm4.9$ & 294.5 & U \\ 
\smallskip
003937.35+152910.7& 1999.78 & 18.61 & 1.68 & 1.05 & 0.30 & 0.17 &      &      & & $25.4\pm4.9$ & 168.4 & D \\ 
004343.87$-$005356.3 &      & 19.60 & 4.33 & 0.97 & 0.29 & 0.17 &      &      & & $10.3\pm5.1$ & 149.3 & U \\ 
004853.30$-$090435.7 & 2000.74 & 19.44 & 4.72 & 1.64 & 0.59 & 0.25 & 0.14 & 0.67 & 2.43 &$115.7\pm6.4$ & 195.4 & D \\ 
010521.01$-$091742.8 &         & 17.54 & 3.14 & 2.00 & 0.82 & 0.45 & 0.72 & 0.32 & 3.20 &  $5.5\pm6.5$ &  26.3 & G \\ 
010717.91$-$091329.5 & 2000.74 & 18.83 & 2.63 & 1.76 & 0.60 & 0.30 & 0.85 & 0.16 & 2.31 &$178.4\pm5.0$ & 108.4 & D \\ 
\smallskip
012028.56$-$083630.9 & 2000.74 & 16.98 & 2.66 & 1.75 & 0.55 & 0.16 & 0.77 & 0.56 & 1.99 &$152.1\pm4.7$ & 107.7 & D \\ 
012101.10$-$001507.4 &         & 19.23 & 4.26 & 1.46 & 0.54 & 0.14 &      &      & & $15.1\pm6.7$ & 166.4 & U$^{5}$ \\
012150.28+011302.8& 1998.72 & 17.00 & 2.73 & 1.80 & 0.48 & 0.23 & 0.86 & 0.45 & 1.90 & $235.0\pm5.0$ & 122.9 & D$^{6}$ \\
012159.74$-$010142.9 &         & 18.07 & 1.08 & 0.49 & 0.20 & 0.08 & 1.31 &$-$0.59 & 1.18 & $16.1\pm5.0$ &  99.8 & U$^{1}$ \\
012321.07$-$004549.9 &         & 17.67 & 1.76 & 1.23 & 0.34 & 0.24 & 0.91 & 0.29 & 1.55 & $12.8\pm5.0$ & 320.2 & G \\ 
\smallskip
012526.74+000448.5& 1998.73 & 19.35 & 1.69 & 1.66 & 0.54 & 0.16 &      &      & & $57.2\pm6.7$ &  96.9 & D \\ 
012747.73$-$100439.2 & 2000.74 & 18.82 & 3.04 & 1.74 & 0.58 & 0.30 & 0.39 & 0.66 & 2.25 & $33.7\pm6.2$ & 211.9 & D \\ 
013007.13+002635.3& 1998.73 & 17.63 & 0.41 & 1.07 & 0.44 & 0.38 & 0.71 & 0.54 & 2.04 & $19.7\pm5.0$ & 131.8 & D$^{7}$ \\
\\
\\
\\
\\
\\
\\
\\
\\
013148.46+004230.7 & 1998.72 & 19.05 & 1.85 & 1.05 & 0.17 & 0.19 &      &      & & $19.1\pm5.0$ &  78.1 & D \\ 
013249.99+000300.4 &         & 20.52 & 1.36 & 1.21 & 0.34 & 0.47 &      &      &   &           &       & ... \\ 
014140.31$-$100355.1 &       & 19.60 & 3.99 & 1.76 & 0.55 & 0.33 &      &      & & $19.2\pm6.6$ & 152.1 & U \\ 
015025.80$-$001315.1 &       & 17.62 & 1.09 & 0.42 & 0.09 & 0.05 & 0.30 &$-$0.31 & 1.10 & $6.4\pm4.8$ & 143.1 & U$^{1}$ \\
\smallskip
015232.31$-$004932.6& 1998.72 & 18.15 & 2.69 & 1.68 & 0.48 & 0.26 & 0.66 & 0.13 & 2.04 & $14.9\pm4.8$ & 128.9 & D \\ 
020716.86+133034.7 &         & 19.49 & 2.04 & 1.02 & 0.28 & 0.28 &      &      & &  $8.7\pm4.6$ & 176.2 & U \\ 
021119.82+003201.5 &         & 19.73 & 3.30 & 1.65 & 0.56 & 0.28 &      &      & &  $6.4\pm6.6$ & 249.2 & U \\
023208.60+003639.3& 1998.73 & 19.46 & 2.09 & 1.59 & 0.56 & 0.28 &      &      & & $21.7\pm6.8$ &  19.2 & D \\ 
025634.62$-$084853.8& 1999.79 & 19.47 & 1.94 & 1.11 & 0.27 & 0.17 &      &      & & $35.2\pm6.8$ & 137.4 & D \\ 
\smallskip
030837.07+005157.1 &         & 20.63 & 1.92 & 1.89 & 0.44 & 0.41 &      &      &          &    &       & ... \\ 
032955.54$-$000354.0 &       & 17.96 & 1.16 & 0.53 & 0.19 & 0.08 & 0.27 & 0.56 & 1.18 &  $6.5\pm4.8$ & 141.7 & U$^{1}$ \\
033704.05$-$001603.0&        & 18.67 & 2.48 & 1.92 & 0.71 & 0.30 & 0.79 & 0.33 & 2.41 & $12.0\pm5.0$ & 116.1 & U \\ 
034705.41$-$063323.9 & 2000.74 & 16.17 & 1.00 & 0.39 & 0.17 & 0.04 & 0.36 &$-$0.02 & 1.01 & $24.0\pm5.1$ & 171.4 & D$^{1}$ \\
073621.29+390725.2& 2000.32 & 18.43 & 3.21 & 1.52 & 0.46 & 0.16 & 0.68 & 0.65 & 1.67 & $55.6\pm4.7$ & 230.1 & D \\ 
\smallskip
073805.73+274148.8 & 2001.97 & 18.46 & 2.49 & 1.57 & 0.55 & 0.21 & 0.78 & 0.10 & 2.19 & $19.1\pm6.3$ & 159.4 & D \\ 
074109.31+243603.0 &         & 19.95 & 2.17 & 1.51 & 0.53 & 0.04 &      &      &            &  &       & ... \\ 
074638.21+400403.5 &         & 19.52 & 3.08 & 1.68 & 0.51 & 0.21 &      &      &        &      &       & ... \\ 
074710.84+251619.6 &         & 18.12 & 1.06 & 0.42 & 0.15 & 0.10 &      &      &  &  $10.3\pm4.5$ & 177.5 & U$^{1}$ \\
075116.37+391201.4&          & 17.61 & 3.47 & 1.60 & 0.58 & 0.43 & 0.81 & 0.30 & 2.43 &   $3.9\pm4.7$ & 168.9 & G \\ 
\smallskip
075228.63+280547.5 & 2002.02 & 20.09 & 1.55 & 1.63 & 0.49 & 0.10 &      &      &   & $34.4\pm6.6$ & 191.2 & D \\ 
075953.64+434021.1&          & 19.48 & 3.64 & 1.91 & 0.63 & 0.15 &      &      &  &   $3.9\pm6.6$ & 324.3 & U \\ 
080040.60+292437.9 &         & 20.46 & 2.13 & 1.78 & 0.58 & 0.06 &      &      &       &      &       & ...$^{8}$ \\ 
\\
\\
\\
\\
\\
\\
\\
\\
080046.72+364107.1 &         & 17.37 & 1.06 & 0.41 & 0.13 & 0.08 & 0.63 &$-$0.01 & 1.04 &   $3.5\pm4.8$ & 220.4 & U$^{1}$ \\
080156.39+395043.8 &         & 18.46 & 4.16 & 1.65 & 0.54 & 0.26 & 0.59 & 0.50 & 2.16 &  $2.0\pm5.0$ & 125.7 & U \\ 
080806.06+293634.1 &         & 18.34 & 3.22 & 1.18 & 0.33 & 0.35 & 0.71 & 0.10 & 1.95 &  $6.3\pm5.0$ &   6.6 & U \\ 
080908.15+360129.0 &         & 18.89 & 0.32 & 0.44 & 0.25 & 0.10 &      &      & &  $6.1\pm4.8$ & 192.4 & U \\ 
\smallskip
082127.90+333037.2 &         & 19.46 & 2.62 & 1.21 & 0.33 & 0.22 &      &      &  & $11.0\pm21.7$& 256.9 & U \\ 
082251.41+461232.3& 2000.26 & 17.23 & 2.79 & 1.40 & 0.47 & 0.23 & 0.82 & 0.20 & 1.73 &  $27.1\pm5.0$ & 221.3 & D \\ 
082626.76+470911.7& 2000.26 & 17.77 & 2.80 & 1.46 & 0.48 & 0.31 & 0.60 & 0.17 & 1.99 &  $39.6\pm4.8$ & 148.9 & D \\ 
084130.12+435136.7 & 2001.15 & 19.45 & 2.52 & 1.52 & 0.48 & 0.23 &      &      & & $33.6\pm6.3$ & 354.7 & D \\ 
084720.16+003340.4 &         & 20.73 & 1.14 & 1.92 & 0.56 & 0.21 &      &   &   &             &       & ... \\ 
\smallskip
085212.77+383721.9 & 2002.02 & 19.93 & 3.10 & 1.69 & 0.51 & 0.18 &      &      & & $51.8\pm6.4$ & 204.2 & D \\ 
085853.28+012243.5& 2000.34 & 18.31 & 2.46 & 1.99 & 0.67 & 0.36 & 0.71 & 0.72 & 2.49 &  $32.1\pm4.7$ & 124.3 & D \\ 
090011.35$-$003606.5& 1999.22 & 18.44 & 3.27 & 1.48 & 0.47 & 0.26 &      &     & &  $53.8\pm6.4$ & 168.4 & D \\ 
090208.05+435503.8 &         & 17.43 & 1.19 & 0.47 & 0.14 & 0.06 &      &      &     &   $4.5\pm4.8$ & 135.2 & U$^{1}$ \\
091007.61+521612.4 &         & 17.01 & 2.94 & 1.63 & 0.54 & 0.27 & 0.72 & 0.20 & 2.09 &  $12.6\pm4.8$ & 177.4 & G \\ 
\smallskip
091019.21+041211.6 &         & 19.05 & 3.42 & 1.50 & 0.54 & 0.38 &      &      & & $12.2\pm6.0$ & 279.3 & U \\ 
091335.63+492444.7 &         & 19.83 & 2.12 & 1.60 & 0.49 & 0.25 &      &      & & $8.0\pm6.5$ & 212.6 & U \\ 
091336.99+034839.1 & 2001.14 & 18.86 & 2.34 & 1.27 & 0.38 & 0.23 &      &      & & $35.4\pm6.0$ & 150.0 & D \\ 
092508.21+010545.4 &         & 17.94 & 2.66 & 1.29 & 0.44 & 0.32 & 0.49 & 0.27 & 2.05 &   $6.3\pm4.8$ & 190.6 & G \\ 
092545.46+424929.0 & 2002.02 & 17.51 & 1.11 & 0.43 & 0.11 & 0.04 & 0.25 &$-$0.26 & 0.98 &  $20.5\pm4.8$ & 156.6 & D$^{1}$ \\
\smallskip
093057.12+033453.6 &         & 19.23 & 2.41 & 1.60 & 0.55 & 0.35 &      &      & & $16.5\pm6.1$ & 349.8 & U \\ 
093203.67+002753.8 &         & 20.40 & 1.00 & 1.66 & 0.45 & 0.10 &      &   &   &             &       & ... \\ 
093741.12+470836.2 & 2001.96 & 18.6\tablenotemark{g}  &      &      &      &      & 0.20 & 0.93 &     &  $61.5\pm4.8$ & 199.9 & D \\ 
094209.94+535719.2 &         & 16.49 & 2.61 & 1.44 & 0.53 & 0.30 & 0.72 & 0.21 & 1.98 &   $7.2\pm4.8$ & 352.4 & G \\ 
\\
\\
\\
\\
\\
\\
\\
094318.73+032743.7 & 2001.14 & 18.11 & 2.70 & 1.38 & 0.49 & 0.13 & 0.54 & 0.04 & 1.66 &  $34.9\pm4.5$ & 229.4 & D \\ 
094511.57+503214.5 & 2002.02 & 18.86 & 5.21 & 1.32 & 0.40 & 0.22 &      &      & &  $15.4\pm4.8$ & 202.7 & D$^{3}$ \\
094858.68+583020.5& 2000.26 & 18.77 & 2.55 & 1.15 & 0.33 & 0.22 &      &   &   &  $26.9\pm4.7$ & 163.2 & D \\ 
095005.09+584123.8 & 2000.26 & 17.03 & 2.93 & 1.43 & 0.47 & 0.18 & 0.67 & 0.41 & 1.76 & $102.9\pm4.7$ & 162.5 & D \\ 
\smallskip
095516.39+012129.8&          & 18.36 & 2.85 & 1.60 & 0.53 & 0.26 & 0.86 &$-$0.05 & 1.97 &   $7.4\pm4.9$ & 100.9 & U \\ 
100132.62+020402.7 &         & 19.41 & 3.84 & 1.68 & 0.61 & 0.20 &      &      & & $10.9\pm6.3$ & 237.9 & U \\ 
100432.45+004337.7& 1999.22 & 19.99 & 1.78 & 1.71 & 0.56 & 0.19 &      &  &    &  $32.1\pm6.4$ & 169.1 & D \\ 
100627.47+462117.4 &         & 17.52 & 0.96 & 0.44 & 0.15 & 0.10 & 0.26 & 0.19 & 1.31 &  $10.4\pm5.0$ & 164.5 & U$^{1}$ \\
100638.20+473039.5 & 2001.97 & 19.96 & 1.88 & 1.65 & 0.54 & 0.27 &      &      & &  $37.2\pm6.7$ & 161.4 & D \\ 
\smallskip
100812.04+012657.2 & 2000.34 & 20.17 & 3.60 & 1.73 & 0.51 & 0.15 &      &   &   &  $425.2\pm6.3$ & 190.7 & D \\ 
100913.95+514924.6 & 2002.02 & 18.55 & 2.66 & 1.44 & 0.45 & 0.16 & 0.64 & 0.15 & 1.80 &  $75.8\pm5.0$ & 217.4 & D \\ 
100958.70+010313.2 & 1999.22 & 17.07 & 2.26 & 1.03 & 0.27 & 0.14 & 0.41 & 0.03 & 1.44 &  $35.9\pm4.8$ & 171.8 & D \\ 
101007.11+031444.1 &         & 18.96 & 1.97 & 1.78 & 0.63 & 0.40 & 1.03 & 0.21 & 2.25 &  $8.9\pm6.6$ & 129.9 & U \\ 
101210.81+465249.1 & 2002.04 & 18.36 & 2.76 & 1.39 & 0.46 & 0.29 & 1.01 & 0.39 & 1.66 & $22.6\pm6.6$ & 218.5 & D \\ 
\smallskip
101422.75+464737.3 &         & 19.77 & 2.34 & 1.52 & 0.52 & 0.09 &      &      & & $18.5\pm6.6$ & 149.1 & U \\ 
102239.58+622344.7 &         & 19.88 & 2.24 & 1.79 & 0.37 & 0.34 &      &      & & $11.2\pm6.4$ & 250.8 & U \\ 
102433.76+562855.1 &         & 19.51 & 2.37 & 1.78 & 0.56 & 0.20 &      &      &  & $3.5\pm6.5$ & 308.6 & U \\ 
102908.38+591853.8 &         & 18.93 & 1.91 & 1.36 & 0.46 & 0.18 &      &      & & $10.9\pm4.9$ & 175.0 & U \\ 
102941.07+020613.1 &         & 19.48 & 2.62 & 1.75 & 0.58 & 0.26 &      &      & & $10.7\pm6.2$ & 169.4 & U \\ 
\smallskip
103758.57+612559.3 & 2000.26 & 18.08 & 1.00 & 0.36 & 0.19 &$-$0.01 &      &      & & $17.9\pm4.9$ & 156.8 & D$^{1}$ \\
104138.55+643219.8 &         & 18.48 & 2.45 & 1.89 & 0.75 & 0.42 & 0.68 & 0.58 & 2.58 &   $3.1\pm6.4$ & 208.4 & U \\ 
104316.17+573755.7 &         & 19.96 & 1.54 & 1.55 & 0.54 & 0.11 &      &      &      &       &       & ... \\ 
\\
\\
\\
\\
\\
\\
\\
\\
105629.95+012208.8 &         & 17.52 & 1.09 & 0.54 & 0.18 & 0.05 & 0.09 &$-$0.77 & 1.32 &   $6.6\pm5.0$ & 180.1 & U$^{1}$ \\
105806.72+032514.7 &         & 18.40 & 2.77 & 1.28 & 0.37 & 0.23 & 0.59 & 0.04 & 1.93 &   $4.0\pm6.3$ & 160.5 & U \\ 
110558.06+510942.9 & 2001.97 & 18.91 & 3.76 & 1.88 & 0.54 & 0.15 & 0.99 & 0.73 & 1.93 &  $51.2\pm6.7$ & 204.3 & D \\ 
110607.96+495055.0 & 2001.97 & 18.82 & 2.89 & 1.56 & 0.55 & 0.14 &      &      & & $65.0\pm6.4$ & 200.0 & D \\ 
\smallskip
111833.84+563958.1 & 2001.89 & 16.65 & 0.88 & 0.35 & 0.15 & 0.02 & 0.27 & 0.18 & 0.98 &  $19.8\pm4.8$ & 222.0 & D$^{1}$ \\
112034.87+555937.2 & 2001.97 & 16.15 & 2.75 & 1.31 & 0.43 & 0.15 & 0.64 & 0.14 & 1.74 &  $95.3\pm4.5$ & 200.9 & D \\ 
112417.34+054921.3 &         & 20.30 & 4.73 & 1.19 & 0.31 & 0.24 &      &      & & $23.4\pm30.7$ & 359.6 & U \\ 
112631.31+053705.3 &         & 20.26 & 3.42 & 1.79 & 0.55 & 0.23 &      &      & & $15.4\pm6.7$ & 166.7 & U \\ 
112650.77+645646.0 &         & 19.09 & 3.15 & 1.39 & 0.45 & 0.25 &      &   &   &  $11.2\pm4.8$ & 198.7 & U \\ 
\smallskip
112654.85+645155.8 &         & 20.30 & 2.27 & 1.78 & 0.57 & 0.09 &      &   &   &   $6.4\pm6.6$ & 225.9 & U \\ 
112746.45+041502.5 & 2001.14 & 18.92 & 2.99 & 1.54 & 0.49 & 0.29 &      &      & & $51.9\pm5.0$ & 148.7 & D \\ 
112752.79+515815.2 & 2001.97 & 17.68 & 2.89 & 1.21 & 0.38 & 0.24 & 0.61 & 0.27 & 1.66 &  $29.1\pm4.5$ & 169.9 & D \\ 
112801.68+004034.7&          & 18.81 & 2.46 & 1.41 & 0.38 & 0.23 &      & &     &  $19.0\pm6.8$ & 171.0 & U$^{9}$ \\
112836.57+011331.2 &         & 17.23 & 1.03 & 0.43 & 0.17 & 0.09 & 0.60 & 0.01 & 1.05 &   $2.2\pm5.2$ & 288.5 & U$^{1}$ \\
\smallskip
112900.75+510742.0 & 2001.97 & 20.14 & 1.81 & 1.44 & 0.48 & 0.04 &      &      & & $23.0\pm6.0$ & 167.2 & D \\ 
112950.38+003344.9& 1999.22 & 18.28 & 2.27 & 1.40 & 0.41 & 0.33 & 0.77 & 0.17 & 1.81 &  $39.6\pm6.8$ & 234.2 & D \\ 
113040.49+525039.2 &        & 17.34 & 1.14 & 0.45 & 0.13 & 0.03 & 0.28 &$-$0.36 & 1.01 &   $6.4\pm4.5$ & 266.8 & U$^{1}$ \\
113720.88$-$012515.7 & 2000.17 & 19.14 & 3.34 & 1.28 & 0.37 & 0.23 &      &      & & $29.6\pm6.2$ & 230.5 & D \\ 
113931.62+050231.2 &         & 19.59 & 2.03 & 1.69 & 0.56 & 0.39 &      &      & & $17.5\pm6.3$ & 164.8 & U \\ 
\smallskip
114125.85+010504.3&         & 17.29 & 2.12 & 0.89 & 0.15 & 0.29 & 0.42 & 0.35 & 1.50 &   $3.9\pm4.8$ &  85.7 & G \\ 
114355.88+625924.7 & 2001.38 & 19.41 & 1.62 & 1.55 & 0.51 & 0.32 &      &      & & $35.2\pm6.5$ & 167.6 & D \\ 
114442.36+043641.9 & 2001.29 & 19.17 & 2.31 & 1.39 & 0.31 & 0.29 &      &      & & $35.0\pm6.2$ & 275.5 & D \\ 
\\
\\
\\
\\
\\
\\
\\
\\
114731.68+003724.5& 1999.22 & 18.60 & 1.71 & 1.05 & 0.23 & 0.14 &      &  &    &  $27.8\pm4.8$ & 171.9 & D \\ 
115057.09$-$012709.7 & 2000.17 & 20.00 & 1.43 & 1.89 & 0.49 & 0.38 &      &  &    &  $23.0\pm6.3$ & 233.6 & D \\ 
115255.58+055337.7 &         & 17.39 & 1.09 & 0.59 & 0.15 & 0.10 & 0.32 & 1.01 & 0.94 &  $10.8\pm4.6$ & 167.1 & U$^{1}$ \\
115555.74+490012.1 & 2002.22 & 18.81 & 2.11 & 1.76 & 0.83 & 0.06 & 0.76 & 0.66 & 2.07 &  $40.3\pm6.7$ & 224.8 & D$^{3}$ \\
\smallskip
115925.70$-$031452.0&        & 18.76 & 1.79 & 1.08 & 0.17 & 0.10 &      &    &  &   $7.7\pm5.0$ &  84.8 & U \\ 
115929.28+640501.8 & 2001.07 & 17.66 & 2.84 & 1.54 & 0.51 & 0.19 & 0.49 & 0.14 & 2.07 & $120.3\pm4.7$ & 233.8 & D \\ 
120031.66+035002.8 &         & 20.14 & 3.38 & 1.72 & 0.48 & 0.24 &      &      &  &           &       & ... \\ 
120241.19+604322.6 &         & 19.85 & 2.57 & 1.43 & 0.44 & 0.12 &      &      &  &           &       & ... \\ 
120336.68+492232.6 & 2002.11 & 19.50 & 2.37 & 1.61 & 0.55 & 0.07 &      &      & & $47.4\pm6.7$ & 253.4 & D \\ 
\smallskip
120941.74$-$023931.7 &        & 18.24 & 2.90 & 1.40 & 0.46 & 0.35 & 0.85 & 0.24 & 1.84 &   $6.9\pm5.0$ & 263.4 & U \\ 
121243.75+005722.3 & 1999.22 & 18.90 & 2.27 & 1.17 & 0.31 & 0.28 &      &   &   &  $40.3\pm6.8$ & 204.2 & D \\ 
121733.35$-$001857.5 &        & 17.18 & 1.22 & 0.55 & 0.24 & 0.06 & 0.46 & 0.08 & 1.16 &   $2.1\pm5.1$ & 322.4 & U$^{1}$ \\
121933.50$-$012843.0 &         & 17.29 & 1.21 & 0.48 & 0.21 & 0.17 & 0.43 & 0.28 & 1.23 &   $2.2\pm5.0$ &  47.9 & U$^{1}$ \\
122843.93+532937.7 & 2001.97 & 19.36 & 3.67 & 1.41 & 0.45 & 0.19 &      &      & & $47.6\pm6.4$ & 171.8 & D \\ 
\smallskip
123626.98+025901.8 &         & 18.98 & 2.62 & 1.79 & 0.55 & 0.40 &      &      & &  $6.8\pm6.8$ & 311.7 & U \\ 
123834.40+023856.0 &         & 18.49 & 0.57 & 1.38 & 0.23 & 0.20 &      &      & &  $8.9\pm5.1$ & 128.8 & U$^{1}$ \\
124114.82+605402.9 &         & 19.94 & 1.55 & 1.61 & 0.56 & 0.05 &      &      & & $74.4\pm32.1$ & 198.2 & U \\ 
124116.94+051942.2 & 2001.21 & 19.48 & 1.55 & 1.24 & 0.36 & 0.17 &      &      & & $28.6\pm6.8$ & 220.4 & D \\ 
124120.87+032848.2 &         & 20.61 & 1.92 & 0.88 & 0.31 & 0.63 &      &      & &            &       & ... \\ 
\smallskip
124215.10+635616.6 & 2001.21 & 19.22 & 3.28 & 1.57 & 0.51 & 0.19 &      &      & & $53.6\pm6.4$ & 171.1 & D \\ 
124436.20+051309.1 &         & 19.49 & 1.95 & 1.65 & 0.52 & 0.31 &      &      & &  $8.4\pm6.0$ & 318.1 & U \\ 
124505.36+044053.3 &         & 18.31 & 2.01 & 0.94 & 0.19 & 0.23 &      &      & &  $2.6\pm4.5$ & 211.1 & U \\ 
\\
\\
\\
\\
\\
\\
\\
\\
125149.87+013001.8 &         & 16.24 & 4.08 & 3.52 & 1.06 & 0.49 & 1.12 & 0.75 & 4.17 &   $6.1\pm5.1$ &  62.2 & G$^{10}$ \\
125431.75+593053.6 &         & 19.41 & 3.24 & 1.41 & 0.48 & 0.29 &      &      & & $13.7\pm6.7$ & 278.4 & U \\ 
125450.29+614842.3 &         & 19.56 & 3.38 & 1.74 & 0.51 & 0.36 &      &      & & $11.6\pm6.7$ & 179.6 & U \\ 
125520.36+631326.7 & 2001.39 & 17.25 & 2.74 & 1.38 & 0.47 & 0.18 & 0.51 & 0.30 & 1.69 &  $94.5\pm49.0$ & 175.2 & G \\ 
\smallskip
130219.89+620559.0 & 2001.39 & 18.66 & 2.09 & 1.00 & 0.25 & 0.07 &      &      & & $28.6\pm4.8$ & 178.2 & D \\ 
130227.53+055512.3 & 2001.29 & 19.25 & 4.59 & 1.85 & 0.55 & 0.25 &      &      & & $27.5\pm6.0$ & 182.7 & D \\ 
130510.57+054844.0 &         & 20.12 & 2.85 & 1.59 & 0.51 & 0.13 &      &      &  &           &       & ... \\ 
130744.53+600903.7 &         & 16.92 & 1.18 & 0.55 & 0.18 & 0.08 & 0.63 &$-$0.26 & 0.99 &             &       & ...$^{1}$ \\
130854.92$-$002742.3 & 1999.22 & 19.44 & 3.44 & 1.79 & 0.59 & 0.33 & 0.86 & 0.77 & 2.33 &  $23.0\pm6.9$ & 248.6 & D \\ 
\smallskip
131841.09+055427.4 &         & 19.89 & 1.79 & 1.88 & 0.67 & 0.32 &      &      & & $17.1\pm6.8$ & 171.4 & U \\ 
131921.25+025218.9 &         & 19.49 & 2.17 & 1.62 & 0.47 & 0.36 &      &      & & $16.5\pm6.8$ & 208.6 & U \\ 
131935.39+585454.0 &         & 20.35 & 2.63 & 1.56 & 0.51 & 0.05 &      &      & &            &       & ... \\ 
132307.04+052048.0 &         & 19.34 & 2.16 & 1.06 & 0.28 & 0.21 &      &      & &            &       & ... \\ 
132309.75+050647.7 & 2001.21 & 18.61 & 2.07 & 1.10 & 0.31 & 0.15 & 0.54 & 0.56 & 1.90 &  $26.7\pm5.1$ & 152.7 & D \\ 
\smallskip
132604.43+605014.0 &         & 19.16 & 2.55 & 1.47 & 0.45 & 0.27 &      &      & &            &       & ... \\ 
132622.21+003743.7 & 1999.22 & 17.22 & 1.02 & 0.51 & 0.19 & 0.07 & 0.21 & 0.22 & 1.39 &  $25.6\pm5.2$ & 245.1 & D$^{1}$ \\
132840.75+002716.9&         & 19.40 & 1.91 & 1.17 & 0.18 & 0.19 &      &      & & $47.0\pm16.0$ & 117.7 & U \\ 
133034.11$-$022430.3 &        & 19.47 & 4.00 & 1.27 & 0.42 & 0.19 &      &      & &  $4.2\pm6.6$ & 265.4 & U \\ 
133034.42+030307.0 &         & 16.33 & 3.04 & 1.07 & 0.33 & 0.28 & 0.67 & 0.23 & 1.77 &   $8.3\pm5.1$ &  66.4 & G \\ 
\smallskip
134745.49+655730.7 &         & 19.01 & 2.13 & 1.75 & 0.56 & 0.27 & 1.11 & 0.28 & 2.00 &   $7.6\pm6.7$ & 150.8 & U \\ 
135057.54+050841.0 & 2001.29 & 15.96 & 2.75 & 1.42 & 0.41 & 0.29 & 0.64 & 0.21 & 1.96 &  $28.1\pm4.5$ & 279.0 & D \\ 
135111.97+043153.0 & 2001.21 & 19.81 & 2.10 & 1.60 & 0.50 & 0.32 &      &      & & $55.8\pm6.0$ & 210.0 & D \\ 
\\
\\
\\
\\
\\
\\
\\
\\
135124.90+040920.2 &         & 18.67 & 2.77 & 1.30 & 0.40 & 0.25 & 0.78 & 0.54 & 1.91 &   $3.3\pm6.0$ & 198.4 & U \\ 
135155.17+052009.1 & 2001.21 & 17.34 & 2.90 & 1.77 & 0.56 & 0.24 & 0.90 & 0.35 & 2.00 &  $22.5\pm4.5$ & 187.7 & D \\ 
135333.01$-$004039.5& 1999.22 & 16.60 & 2.68 & 1.59 & 0.54 & 0.27 & 0.81 & 0.19 & 1.99 &  $57.0\pm4.8$ & 236.6 & D \\ 
135534.06+021151.6 &         & 20.67 & 1.08 & 1.66 & 0.32 &$-$0.09 &      &      & &            &       & ... \\ 
\smallskip
140327.81+024725.1 & 2000.34 & 19.42 & 1.73 & 1.43 & 0.49 & 0.40 &      &      & & $55.0\pm6.0$ & 149.9 & D \\ 
140411.94$-$014951.4 & 2001.39 & 20.04 & 3.14 & 1.21 & 0.45 &$-$0.10 &      &      & & $22.4\pm6.7$ & 172.5 & D \\ 
140425.25+013252.7 &         & 15.93 & 4.33 & 1.53 & 0.56 & 0.43 & 0.66 & 0.26 & 2.40 &   $2.3\pm5.1$ &  27.6 & G \\ 
141046.22$-$011332.2 & 2001.39 & 19.28 & 1.66 & 1.46 & 0.46 & 0.12 &      &      & & $46.8\pm6.8$ & 276.5 & D \\ 
141111.54+614933.7 & 2001.14 & 18.71 & 2.39 & 1.09 & 0.30 & 0.20 &      &      & & $28.7\pm4.8$ & 251.8 & D \\ 
\smallskip
141324.15+015547.8 & 2000.34 & 18.73 & 3.02 & 1.93 & 0.52 & 0.28 & 0.58 & 0.54 & 2.33 &  $31.0\pm5.0$ & 290.4 & D \\ 
141412.57+030647.6 &         & 15.92 & 3.08 & 1.18 & 0.35 & 0.32 & 0.67 & 0.13 & 1.92 &   $4.4\pm4.6$ & 310.7 & G \\ 
141421.39+043129.3 &         & 20.22 & 1.33 & 1.78 & 0.49 & 0.24 &      &      & & $65.9\pm26.4$ & 270.1 & U \\ 
142112.42$-$004822.6&         & 19.05 & 2.20 & 1.51 & 0.36 & 0.42 & 1.01 & 0.22 & 2.08 &  $14.5\pm6.8$ &  97.7 & U \\ 
142312.35$-$022824.2 &       & 17.69 & 2.77 & 1.13 & 0.33 & 0.27 & 0.51 & 0.03 & 1.89 &   $7.6\pm5.0$ &  73.5 & G \\ 
\smallskip
142535.12+024730.8 &         & 19.19 & 2.66 & 0.98 & 0.20 & 0.25 &      &   &   &   $4.0\pm6.6$ & 237.5 & U \\ 
143156.34+034228.4 &         & 20.42 & 3.45 & 1.82 & 0.51 & 0.20 &      &   &   &             &       & ... \\ 
143328.12+595808.9 & 2001.14 & 18.02 & 0.96 & 0.44 & 0.18 & 0.08 & 0.58 & 0.05 & 1.55 &  $22.5\pm5.0$ & 171.0 & D$^{1}$ \\
144104.98$-$013824.4 & 2001.39 & 19.44 & 1.62 & 1.85 & 0.60 & 0.19 & 0.52 & 0.96 & 2.48 &  $20.3\pm6.7$ & 153.2 & D \\ 
144150.90$-$002424.4&        & 17.86 & 2.59 & 1.72 & 0.66 & 0.26 & 0.88 & 0.51 & 2.21 &   $3.9\pm5.2$ & 304.6 & U \\ 
\smallskip
144428.79+021430.7 &         & 19.36 & 3.45 & 1.60 & 0.54 & 0.36 &      &      & &  $8.4\pm7.0$ & 310.7 & U \\ 
144448.40+043944.2 & 2000.36 & 19.59 & 3.65 & 1.85 & 0.49 & 0.20 &      &      & & $116.9\pm6.7$ & 291.2 & D \\ 
144841.80+042844.5 &         & 17.19 & 2.30 & 1.11 & 0.30 & 0.24 & 0.49 & 0.13 & 1.64 &   $5.6\pm5.0$ & 291.8 & G \\ 
\\
\\
\\
\\
\\
\\
\\
\\
144845.39+031059.4 &         & 19.21 & 2.15 & 0.97 & 0.23 & 0.22 &      &      & &  $0.9\pm5.0$ &  94.5 & U \\ 
144923.37$-$011732.9 &       & 20.43 & 4.38 & 1.23 & 0.32 & 0.36 &      &      &      &       &       & ... \\ 
144945.37+012656.2 &         & 15.64 & 3.45 & 1.26 & 0.39 & 0.41 & 0.75 & 0.18 & 2.11 &   $1.9\pm5.3$ &  52.3 & G \\ 
144958.04$-$010308.6 & 2001.45 & 20.09 & 1.02 & 1.47 & 0.41 & 0.09 &      &      & & $45.1\pm7.2$ & 256.6 & D \\ 
\smallskip
145318.82+600421.1 & 2000.26 & 17.59 & 3.12 & 1.61 & 0.60 & 0.29 & 0.56 & 0.29 & 1.99 & $232.5\pm5.0$ & 267.7 & D \\ 
145459.75+563807.2 & 2001.38 & 17.85 & 2.64 & 1.61 & 0.47 & 0.24 & 0.69 & 0.53 & 1.96 &  $74.0\pm5.0$ & 179.7 & D \\ 
145506.01+002239.1 &         & 17.71 & 1.15 & 0.48 & 0.21 & 0.06 &      & &     &   $6.5\pm5.4$ & 198.4 & U$^{1}$ \\
145512.68+614101.1 &        & 19.43 & 1.72 & 1.89 & 0.63 & 0.29 & 0.80 & 0.45 & 2.57 &  $10.7\pm6.7$ & 270.6 & U \\ 
145933.45+033310.2 & 2000.36 & 19.02 & 2.30 & 1.33 & 0.43 & 0.29 &      &      & & $26.8\pm6.6$ & 191.1 & D \\ 
\smallskip
150051.84+022656.1 &         & 18.82 & 3.21 & 1.12 & 0.30 & 0.27 &      &      & & $13.5\pm6.6$ & 174.3 & U \\ 
150245.71$-$022454.8 & 2001.39 & 17.86 & 2.79 & 1.34 & 0.38 & 0.22 & 0.10 & 0.44 & 1.86 &  $29.4\pm5.3$ & 230.3 & D \\ 
150430.04+561603.8 &         & 19.15 & 2.08 & 1.14 & 0.33 & 0.17 &      &      & & $14.0\pm5.0$ & 251.6 & U \\ 
150713.81+595402.6 & 2001.22 & 19.36 & 3.24 & 1.42 & 0.44 & 0.37 &      &      & & $67.0\pm4.8$ & 264.0 & D \\ 
152132.17$-$012010.3 &       & 19.20 & 3.60 & 1.74 & 0.50 & 0.29 &      &      & &            &       & ... \\ 
\smallskip
152218.79+582629.3 &         & 19.67 & 2.39 & 1.66 & 0.52 & 0.27 &      &      &    &         &       & ... \\ 
152352.14+460216.5 & 2002.35 & 19.26 & 2.36 & 1.31 & 0.37 & 0.26 &      &      & & $22.0\pm4.8$ & 144.6 & D \\ 
152410.66+453657.3 &         & 20.00 & 1.49 & 1.86 & 0.62 & 0.28 &      &      & & $18.9\pm6.4$ & 217.0 & U \\ 
152434.12+444956.3 & 2002.35 & 19.33 & 2.22 & 1.74 & 0.55 & 0.23 &      &      & & $26.3\pm4.8$ & 187.6 & D \\ 
152525.47+450706.5 &         & 19.54 & 1.39 & 0.68 & 0.29 & 0.11 &      &      & & $14.3\pm4.8$ & 199.3 & U$^{11}$ \\
\smallskip
152702.75+434517.4 & 2002.35 & 16.25 & 2.99 & 1.68 & 0.62 & 0.37 & 0.69 & 0.29 & 2.25 &  $42.9\pm5.0$ & 289.5 & D \\ 
153107.09+522200.6 & 2001.39 & 18.24 & 2.31 & 1.20 & 0.36 & 0.23 & 1.03 & 0.12 & 1.49 &  $15.5\pm5.0$ & 159.5 & D \\ 
153532.93+011016.4 &         & 19.29 & 2.17 & 1.96 & 0.71 & 0.31 & 0.58 & 0.55 & 3.01 &  $11.6\pm6.0$ & 265.0 & U \\ 
\\
\\
\\
\\
\\
\\
\\
\\
153732.19+004343.1& 1999.22 & 17.63 & 2.55 & 1.82 & 0.63 & 0.38 & 0.84 & 0.27 & 2.42 &  $33.6\pm4.6$ & 144.3 & D \\ 
154156.85+514421.3 & 2001.38 & 17.5\tablenotemark{h} &      &      &      &      & 0.50 & 0.26 &  &  $60.5\pm5.1$ & 199.7 & D$^{2}$ \\
154426.04+023623.7 & 2001.21 & 15.67 & 2.73 & 1.40 & 0.44 & 0.30 & 0.71 & 0.15 & 1.92 &  $60.1\pm4.9$ & 221.4 & D \\ 
154839.88+411621.1 & 2002.35 & 19.54 & 2.78 & 1.59 & 0.46 & 0.28 &      &      & & $46.4\pm5.1$ & 251.3 & D \\ 
\smallskip
154900.25+514156.0 & 2001.29 & 17.68 & 3.47 & 1.68 & 0.57 & 0.31 & 0.76 & 0.22 & 2.12 &  $16.1\pm5.0$ & 222.6 & D \\ 
155043.84+571342.5 &         & 20.48 & 1.48 & 1.75 & 0.55 & 0.15 &      &      & &            &       & ... \\ 
155441.23+514601.1 &         & 20.81 & 0.83 & 1.83 & 0.49 & 0.16 &      &      &        &     &       & ... \\ 
155745.94+480207.5 & 2001.38 & 19.15 & 2.16 & 1.52 & 0.47 & 0.26 &      &      & & $27.1\pm5.0$ & 260.2 & D \\ 
161657.55$-$010349.9&        & 19.06 & 2.96 & 1.70 & 0.60 & 0.22 &      &  & & $19.4\pm6.6$ & 174.2 & U$^{3}$ \\
\smallskip
162923.70+455729.6 & 2000.26 & 19.67 & 2.62 & 1.68 & 0.50 & 0.16 &      &      & & $50.1\pm6.7$ & 234.2 & D \\ 
163045.02+004025.2 & 1999.22 & 19.99 & 3.24 & 1.80 & 0.57 & 0.15 &      &   &   &  $57.5\pm6.2$ & 245.0 & D \\ 
164410.81+395642.5 & 2000.26 & 18.73 & 2.70 & 1.70 & 0.52 & 0.22 & 0.21 & 0.87 & 2.19 &  $34.8\pm6.4$ & 256.9 & D \\ 
164619.36+383333.6 & 2001.38 & 19.63 & 2.62 & 1.46 & 0.37 & 0.25 &      &      & & $31.7\pm6.6$ & 238.6 & D \\ 
164650.57+405509.4 &         & 19.53 & 2.73 & 1.58 & 0.48 & 0.28 &      &   &   &   $1.5\pm6.4$ & 222.1 & U \\ 
\smallskip
165141.94+352012.6 &         & 16.65 & 0.89 & 0.36 & 0.14 & 0.06 & 0.40 & 0.03 & 0.95 &   $6.6\pm5.0$ & 203.8 & U$^{1}$ \\
165532.15+410112.6 &         & 20.10 & 2.32 & 1.64 & 0.51 & 0.19 &      & &     &             &       & ... \\ 
165948.02+391241.7 &         & 19.63 & 3.33 & 1.86 & 0.58 & 0.38 &      &  &    &   $8.7\pm5.0$ & 165.7 & U \\ 
170219.09+351034.5 & 2001.39 & 16.01 & 2.39 & 1.07 & 0.28 & 0.19 & 0.56 & 0.16 & 1.55 &  $23.1\pm5.0$ & 142.3 & D \\ 
171303.09+285902.2 & 2001.39 & 18.71 & 2.97 & 1.68 & 0.63 & 0.35 & 0.74 & 0.24 & 2.31 & $22.7\pm4.5$ & 341.5 & D \\ 
\smallskip
171502.33+313231.5 &         & 19.37 & 3.49 & 1.55 & 0.47 & 0.21 &      &      & & $14.1\pm6.1$ & 174.8 & U \\ 
171606.44+275255.3 & 2001.39 & 19.40 & 2.74 & 1.58 & 0.50 & 0.24 &      &      & & $21.6\pm6.1$ & 178.8 & D \\ 
171638.50+274937.7 & 2001.39 & 17.95 & 2.35 & 1.25 & 0.39 & 0.26 & 0.62 & 0.14 & 1.82 &  $38.9\pm4.5$ & 255.6 & D \\ 
171942.38+575837.7&         & 16.81 & 3.29 & 1.18 & 0.33 & 0.34 & 0.62 & 0.09 & 2.04 &  $13.3\pm5.0$ & 222.2 & G$^{12}$ \\
\\
\\
\\
\\
\\
\\
\\
171957.65+575005.5&         & 16.47 & 1.25 & 1.25 & 0.28 & 0.41 & 0.67 & 0.25 & 2.08 &   $4.4\pm5.0$ & 154.7 & G$^{13}$ \\
172038.83+575934.4&         & 17.76 & 2.66 & 1.17 & 0.29 & 0.24 & 0.42 &$-$0.17 & 1.85 &   $5.7\pm5.0$ & 183.9 & G$^{14}$ \\
172053.90+293416.3 &        & 20.15 & 1.97 & 1.79 & 0.59 & 0.26 &      &      & & $16.1\pm6.1$ & 192.0 & U \\ 
\smallskip
172823.74+634507.4 &        & 18.73 & 3.03 & 1.06 & 0.30 & 0.15 &      &   &   &  $10.4\pm4.8$ & 172.2 & U$^{3}$ \\
172909.13+594034.8&         & 20.19 & 0.16 & 1.85 & 0.24 & 0.11 &      &    &  &   0.0       &       & U$^{15}$ \\
173650.55+563800.5&         & 19.19 & 2.53 & 1.73 & 0.67 & 0.32 &      &    &  &   $2.8\pm6.6$ & 276.4 & U \\ 
213419.39$-$063423.8 &      & 19.83 & 1.40 & 1.19 & 0.13 & 0.63 &      &  &    &             &       & ... \\ 
213430.27+003119.1 & 2001.72 & 18.97 & 1.52 & 1.80 & 0.59 & 0.41 & 0.68 & 0.30 & 2.43 &  $21.5\pm6.7$ & 201.9 & D \\ 
\smallskip
213600.87$-$004534.1 &      & 18.46 & 2.78 & 1.59 & 0.55 & 0.22 & 0.66 & 1.04 & 1.94  &          &       & ... \\ 
214650.96+005829.4 & 2001.72 & 18.73 & 2.17 & 1.32 & 0.39 & 0.33 &$-$0.76 & 1.87 & 1.90 &  $40.9\pm6.7$ & 181.1 & D \\ 
215000.91+115343.8 & 2000.74 & 18.66 & 2.60 & 1.71 & 0.59 & 0.27 & 0.74 & 0.41 & 2.19 &  $32.1\pm6.4$ & 262.1 & D \\ 
215003.17$-$005235.6 &       & 16.30 & 2.82 & 1.03 & 0.30 & 0.27 & 0.57 & 0.13 & 1.82 &   $3.9\pm5.1$ & 175.8 & G \\ 
215018.14$-$081339.3 &       & 16.81 & 1.69 & 0.66 & 0.00 & 0.35 & 0.44 &$-$0.01 & 1.21 &   $8.2\pm4.9$ &  75.4 & U$^{1}$ \\
\smallskip
215944.74$-$010630.2 & 1999.79 & 19.25 & 2.47 & 1.44 & 0.46 & 0.25 &      &      & & $26.5\pm5.1$ &  98.8 & D \\ 
221450.93+011250.3& 1999.78 & 19.87 & 2.57 & 1.62 & 0.56 & 0.11 &      &      & & $31.1\pm6.8$ & 215.7 & D \\ 
221506.15$-$090147.9 &        & 18.86 & 3.04 & 1.67 & 0.55 & 0.30 & 0.64 & 0.56 & 2.29 &   $8.6\pm4.8$ & 102.4 & U \\ 
221854.26+010026.1&         & 19.22 & 2.21 & 1.21 & 0.37 & 0.18 &      &      & & $16.4\pm6.7$ & 166.3 & U \\ 
223250.78$-$003436.1 & 1998.72 & 19.43 & 3.95 & 1.87 & 0.57 & 0.40 &      &      & & $185.1\pm4.8$ & 179.5 & D \\ 
\smallskip
223302.74$-$001715.5 &         & 19.36 & 5.55 & 1.93 & 0.60 & 0.15 &      &      & & $13.3\pm6.3$ & 160.2 & U \\ 
223728.04+124213.8 &         & 19.77 & 1.89 & 1.61 & 0.51 & 0.20 &      &      &    &         &       & ... \\ 
223937.82+130924.9 & 2000.74 & 17.53 & 2.66 & 1.53 & 0.34 & 0.23 & 0.75 &$-$0.11 & 1.85 &  $57.4\pm6.5$ &  82.6 & D \\ 
224146.80$-$083243.0 &         & 20.16 & 2.60 & 1.48 & 0.30 & 0.13 &      &      & &            &       & ... \\ 
225723.00$-$085055.7 & 2001.72 & 19.64 & 4.33 & 1.79 & 0.76 & 0.48 &      &      & & $33.4\pm6.4$ & 252.9 & D \\ 
230255.00+005904.4& 1999.79 & 17.69 & 2.42 & 1.57 & 0.54 & 0.32 & 0.81 & 0.20 & 2.07 &  $38.5\pm5.0$ & 231.5 & D \\ 
230954.67$-$093736.6 &         & 18.89 & 2.45 & 1.23 & 0.37 & 0.24 & 0.67 &$-$0.57 & 2.13 &   $7.6\pm5.0$ & 106.0 & U \\ 
\\
\\
\\
\\
\\
231340.54+143600.8 &  2001.72 & 19.7\tablenotemark{h} &     &      &    &      &      &      & & $15.8\pm4.7$ & 169.4 & D$^{2}$ \\
231941.81$-$105318.3 & 2000.74 & 19.84 & 2.62 & 1.80 & 0.48 & 0.23 &      &  &    &  $65.0\pm6.6$ & 183.2 & D \\ 
232310.27+010826.4 &         & 20.22 & 0.94 & 1.66 & 0.44 & 0.24 &      &      &    &         &       & ... \\ 
234038.86+150841.8 &         & 18.92 & 2.66 & 1.51 & 0.50 & 0.29 &      &      &  & $6.3\pm4.6$ & 357.7 & U \\

\enddata

\tablenotetext{a}{~~as per normal convention, coordinate names are truncated
rather than rounded; precise astrometry is available in the SDSS archive}
\tablenotetext{b}{~~Epochs provided for those objects with proper motions 
                  detected with $\ge3\sigma$ significance}
\tablenotetext{c}{~~An approximate transformation to the Cousins I
magnitude, derived empirically from a comparison of SDSS observations
of multiple standard stars with published broadband photometry
\citep{gre04}, is I$_{\rm c} = -0.333 {\it (r-i)} + {\it i} -0.443$.
Note that this transformation is not optimized for the specific case
of C stars.}
\tablenotetext{d}{~~the errors are $\pm1\sigma$ uncertainties}
\tablenotetext{e}{~~Object notes: 
1. ``F/G Carbon star'';
2. possible ``F/G Carbon star'';
3. poorly calibrated spectrum;
4. FASST 2; see \citet{hs98};
5. RASS and FIRST source nearby;
6. LP 587$-$45;
7. composite;
8. variable?;
9. candidate extragalactic object (see Paper~I);
10. N-type; see \citet{ti98} and \citet{hs98};
11. RASS source (QSO) nearby;
12. in Draco dwarf galaxy (BASV 461);
13. in Draco dwarf galaxy (Draco C-1); symbiotic variable; ROSAT source;
14. in Draco dwarf galaxy (BASV 578);
15. in Draco dwarf galaxy?}
\tablenotetext{f}{~~D = Dwarf, G = Giant, U = Uncertain, ... = no data}
\tablenotetext{g}{~~poor photometry: {\it g} mag given for crude guidance}
\tablenotetext{h}{~~poor photometry}

\end{deluxetable}

\clearpage

\begin{deluxetable}{rccccccl}
\footnotesize
\tablenum{2}
\tablecaption{Variable Carbon Stars\label{tbl-2}}
\tablewidth{0pt}
\tablehead{
\colhead{Name SDSS J+} & \colhead{Luminosity} & \colhead{{\it u}} & \colhead{{\it g}}&
\colhead{{\it r}} &\colhead{{\it i}} &\colhead{{\it z}} &
\colhead{Date} 
}

\startdata

SDSS J080040.60+292437.9  &  uncertain  &   24.380&22.245&20.463&19.887&19.830& 19 Jan. 2002 \\
                          &             &   23.362&22.221&20.679&19.993&19.755& 21 Nov. 2001 \\
\enddata
\end{deluxetable}

\clearpage

\begin{figure}
\epsscale{0.8}
\plotone{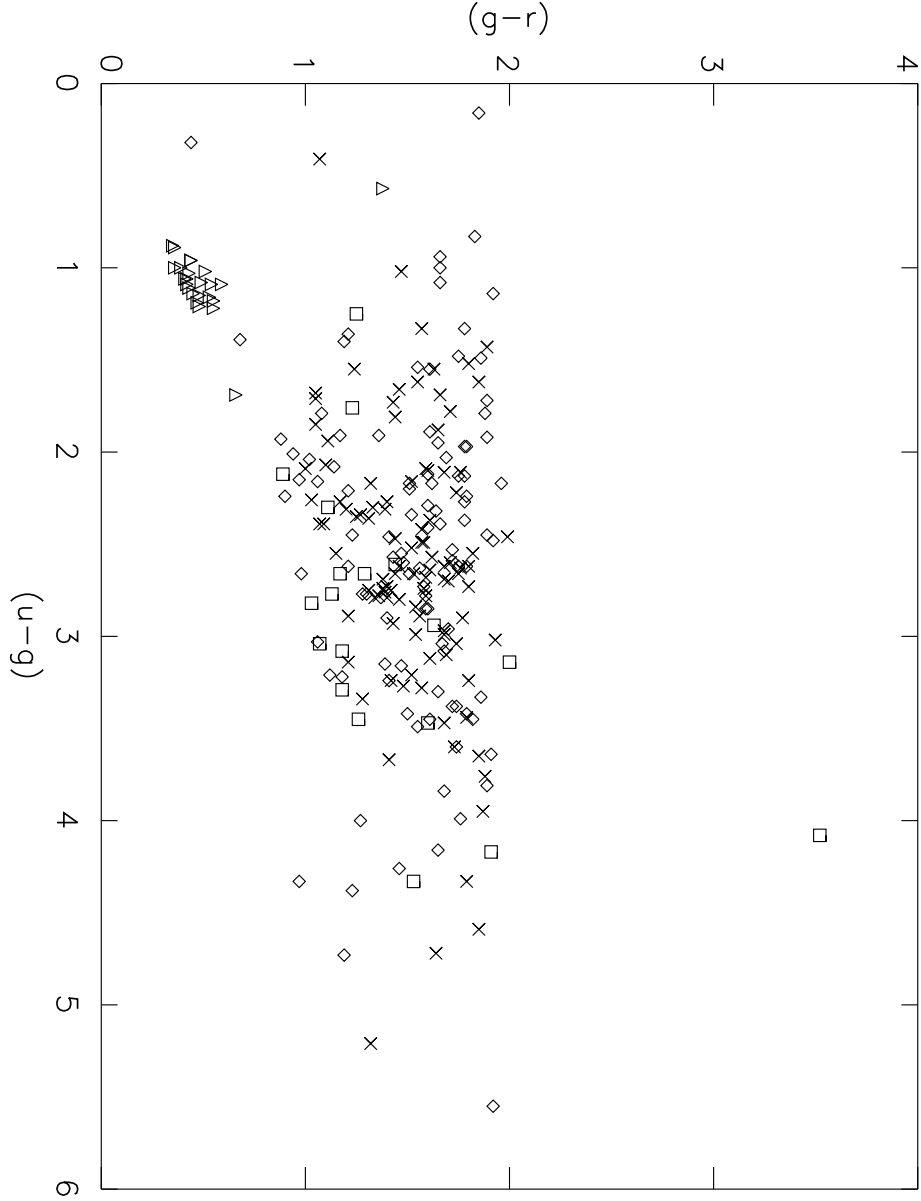}
\caption[fig1.ps]{{\it (u-g)} vs. {\it (g-r)} diagram for the SDSS FHLCs. 
Dwarfs are indicated with an ``X'', giants with a square, ``F/G Carbon stars''
with a triangle, and objects with uncertain or unknown luminosities
with a diamond.  Note the one SDSS N-type star (the object with 
{\it (g-r)} = 3.52).  The ``F/G Carbon stars'' are mostly clumped around
{\it (u-g)} = 1.0, {\it (g-r)} = 0.5.  Other than these rare objects, these
colors provide no effective segregation of carbon giants and dwarfs.
Obviously the {\it u} measurements for the faintest, reddest objects are
highly uncertain.
\label{fig1}}

\end{figure}

\begin{figure}
\epsscale{0.9}
\plotone{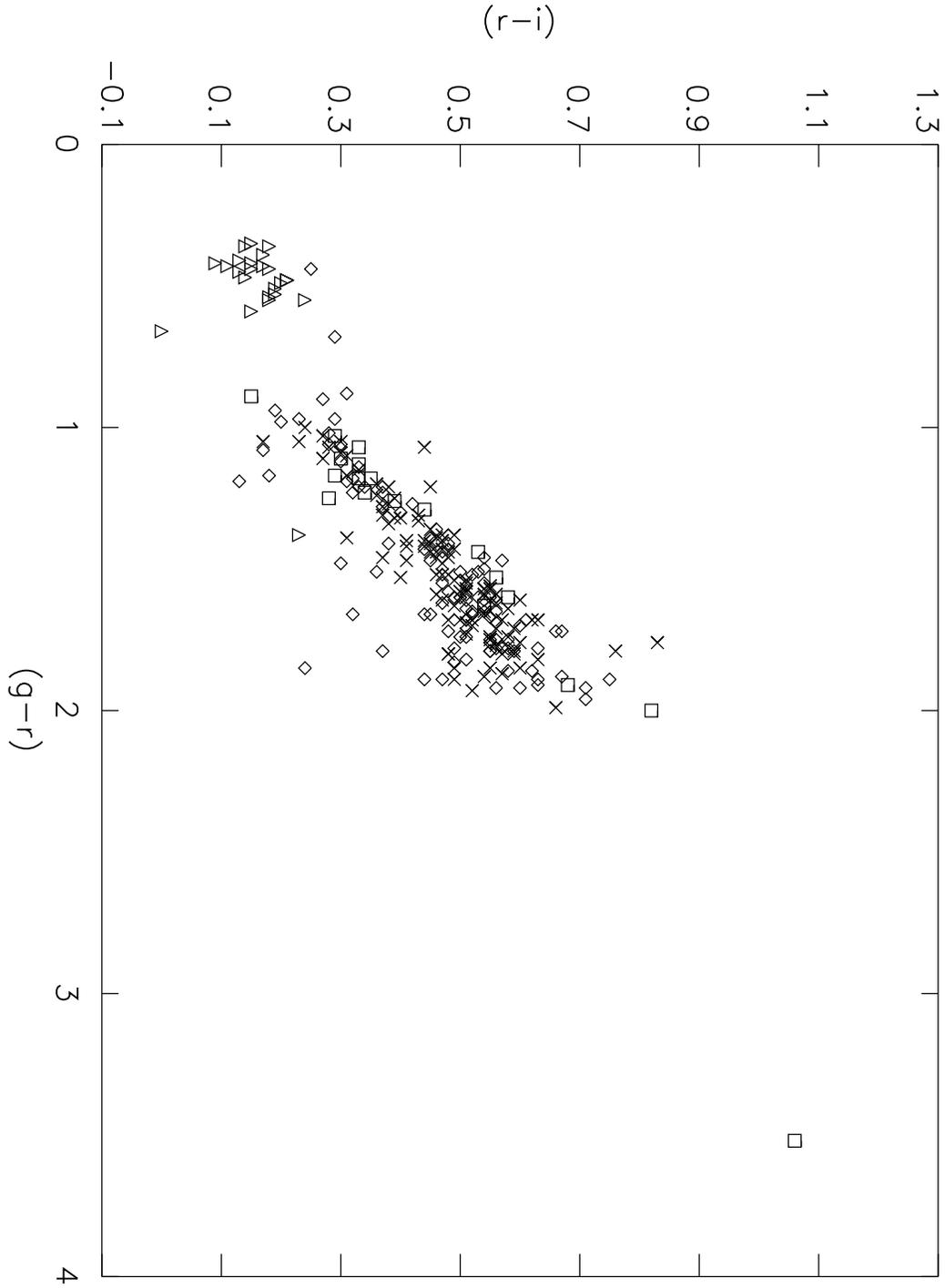}
\caption[fig2.ps]{{\it (g-r)} vs. {\it (r-i)} diagram. Same symbols as in 
Figure~\ref{fig1}.  Note the tight correlation, with the N-type star and the
``F/G Carbon stars'' clearly separate from the majority of the FHLCs (R-type 
stars). However, the small number of the latter objects in our sample makes the
significance of the gap in the color distribution unclear.
\label{fig2}}
\end{figure}

\begin{figure}
\plotone{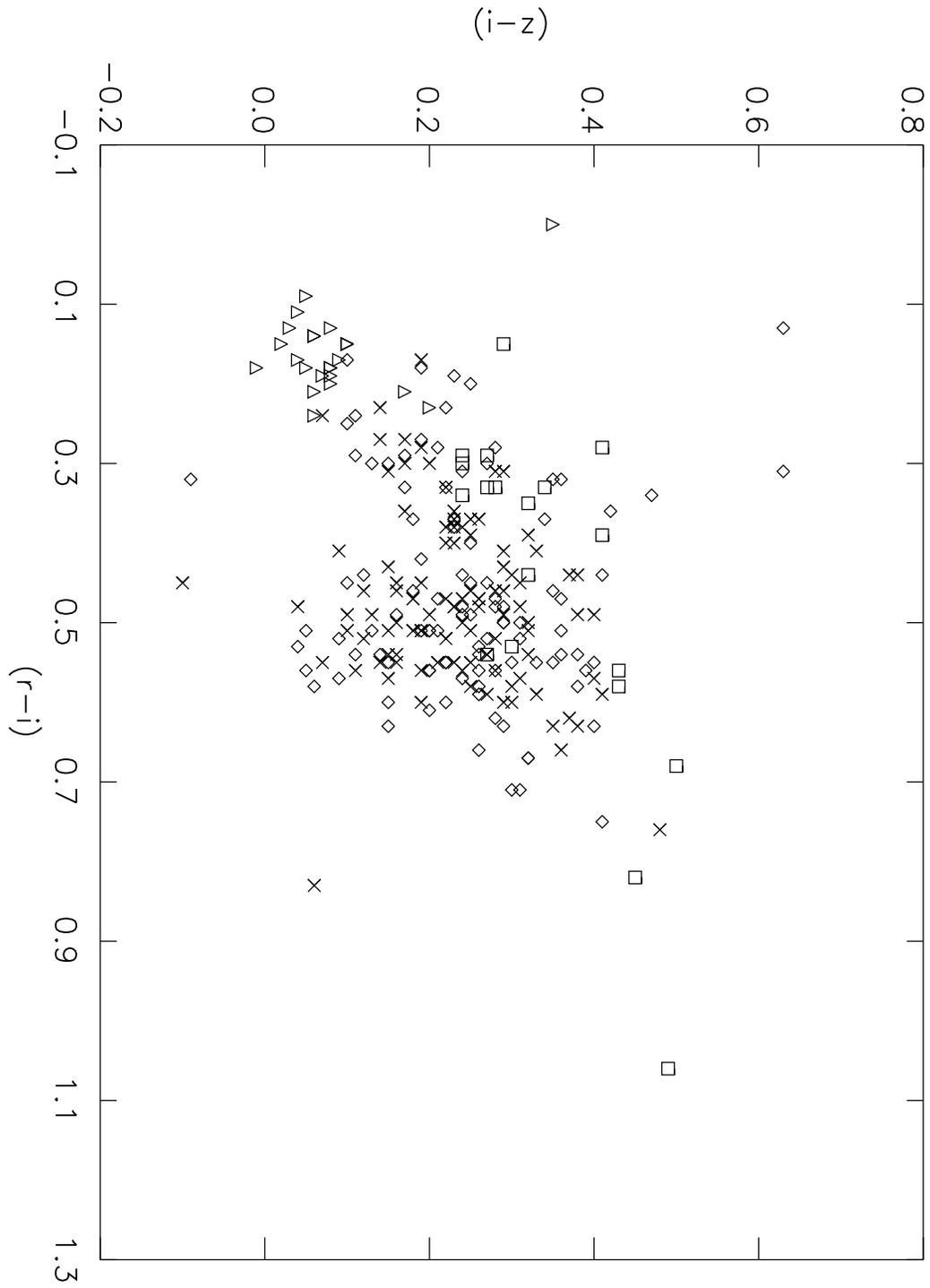}
\caption[fig3.ps]{{\it (r-i)} vs. {\it (i-z)} diagram. Same symbols as in 
Figure~\ref{fig1}.  The N-type (the object with {\it (r-i)} = 1.06), and the
``F/G Carbon stars'' are not nearly as distinct in this diagram.
\label{fig3}}
\end{figure}

\begin{figure}
\plotone{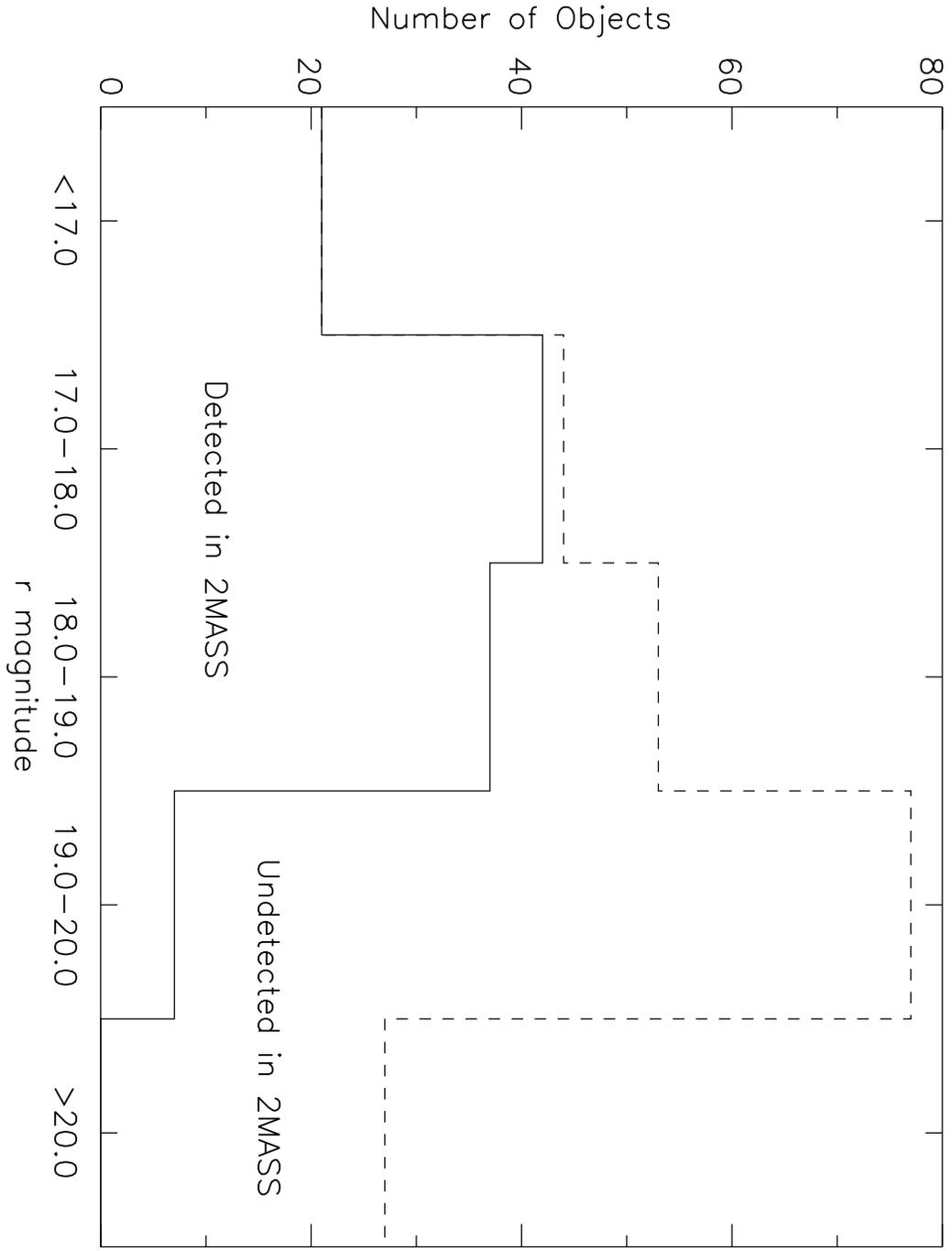}
\caption[fig4.ps]{Comparison of SDSS FHLCs with 2MASS survey.  As expected,
the brighter objects are detected by 2MASS, while the fainter objects are not.
\label{fig4}}
\end{figure}

\begin{figure}
\epsscale{0.8}
\plotone{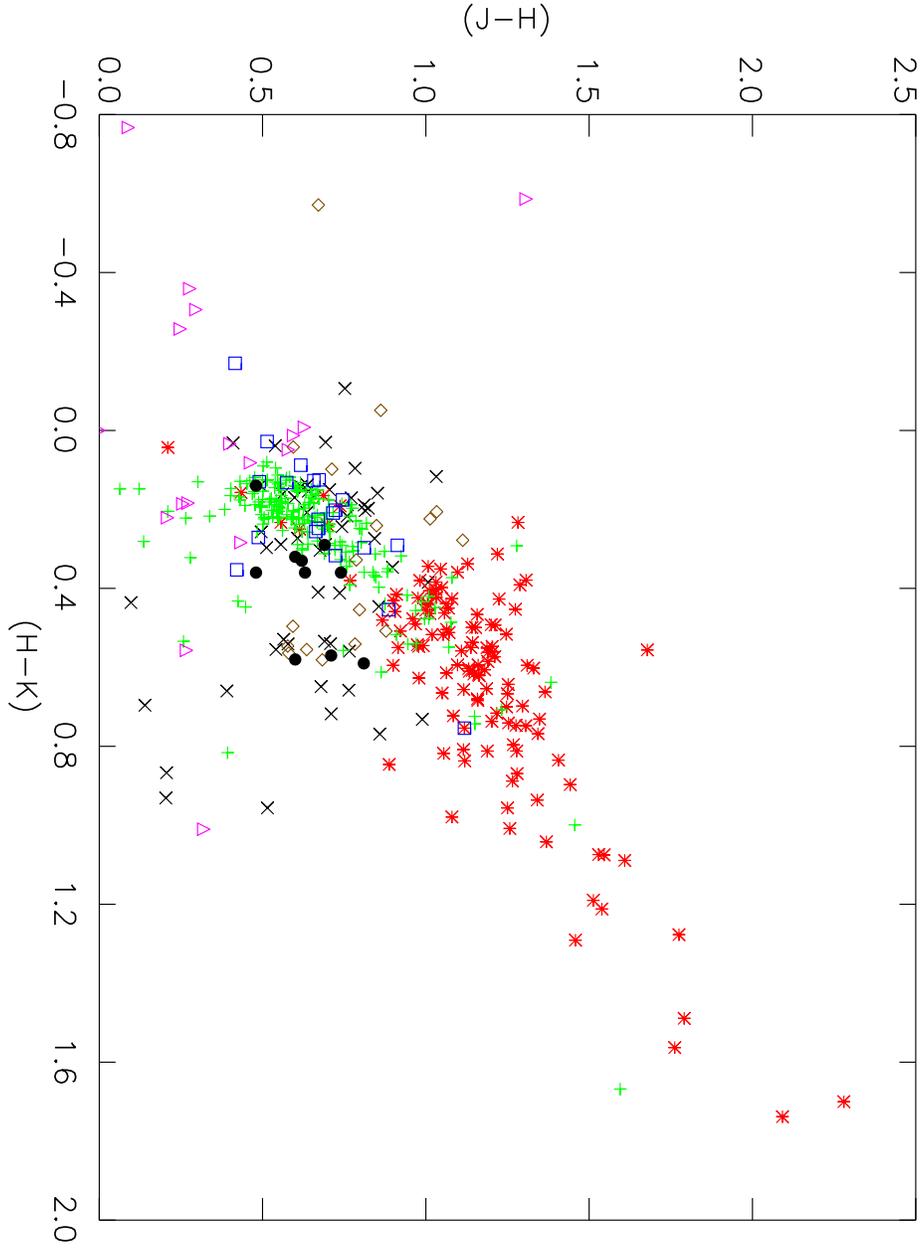}
\caption[fig5.ps]{Near-IR colors for various carbon star samples.
Objects from \citet{alk01} are indicated with a red asterisk (N-type)
and a green plus sign (R-type).  SDSS FHLCs are indicated with a black
``X'' (dwarfs), a blue square (giants), a purple triangle (``F/G Carbon
stars''), and a brown diamond (objects with uncertain/unknown
luminosities).  Previously reported dwarfs are indicated by a filled
black circle.
\label{fig5}}
\end{figure}

\begin{figure}
\plotone{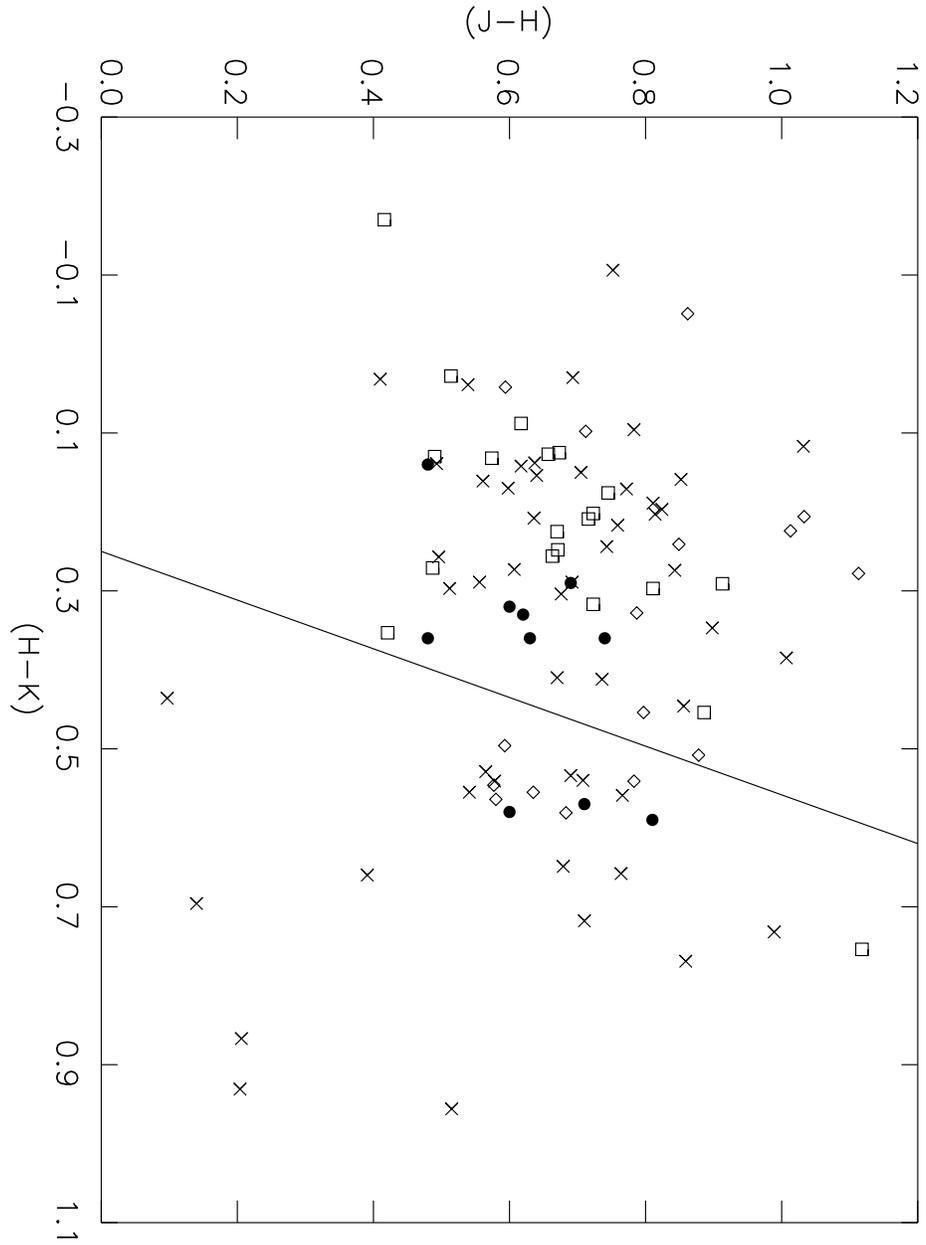}
\caption[fig6.ps]{Expanded near-IR color-color diagram.  Same symbols as
in Figure~\ref{fig5}.  Note that most of the objects to the right of the
solid line are dwarfs.
\label{fig6}}
\end{figure}

\begin{figure}
\epsscale{0.8}
\plotone{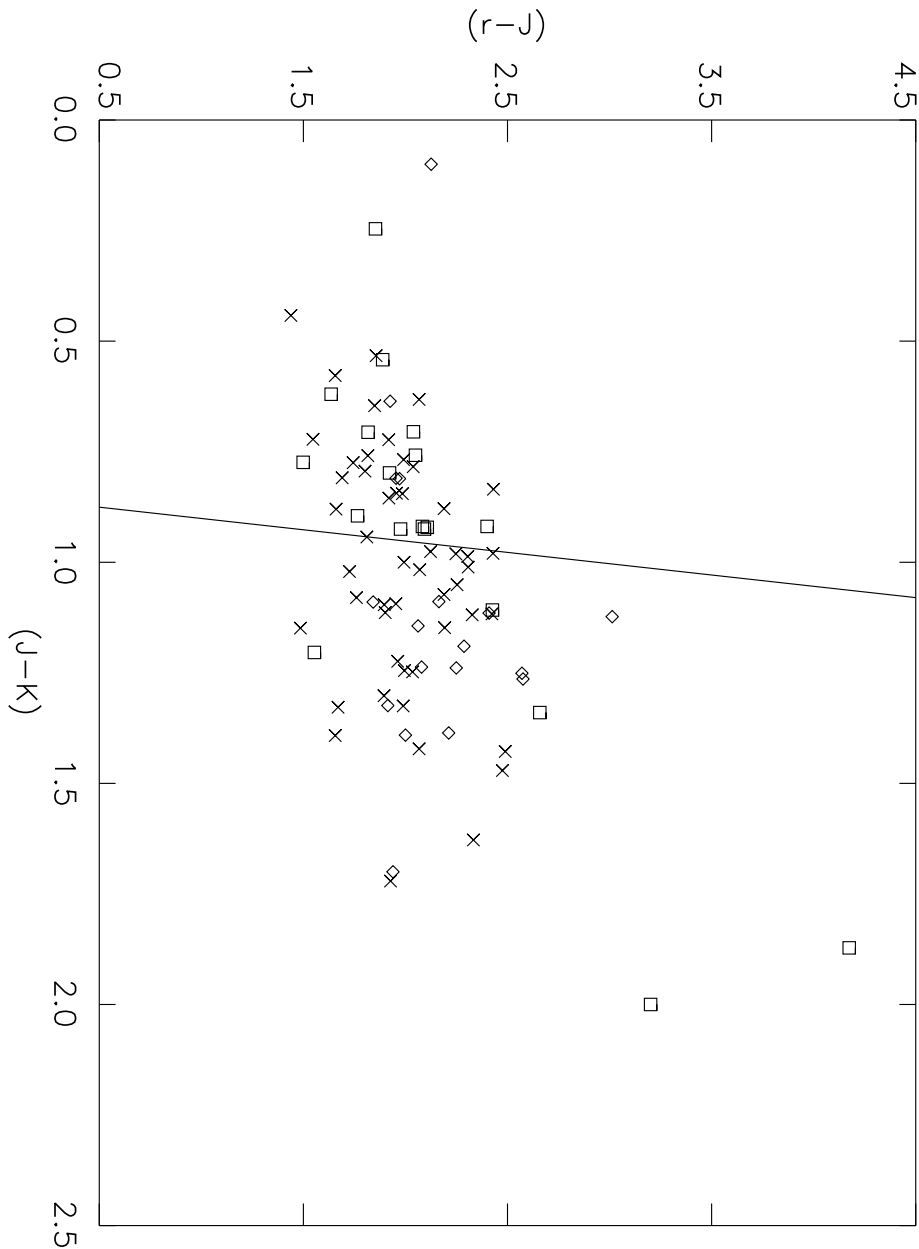}
\caption[fig7.ps]{Photometric luminosity discriminant proposed by
\citet{lkr03}. Same symbols as in Figure~\ref{fig5}, and the solid
line is the luminosity segregation criterion proposed by those
authors.  This indicator is not as effective as that using the near-IR
colors.
\label{fig7}}
\end{figure}

\begin{figure}
\epsscale{0.8}
\plotone{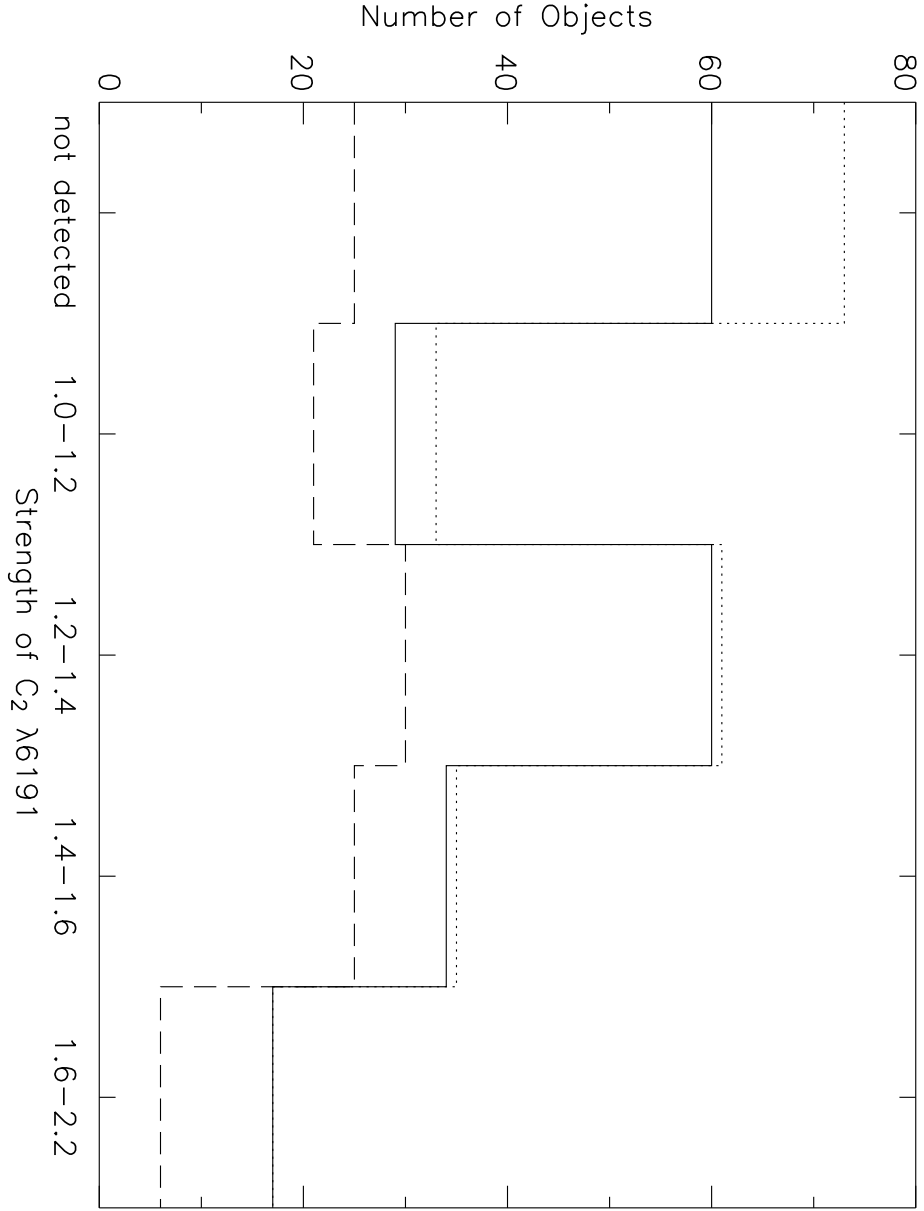}
\caption[fig8.ps]{Histogram of C$_{2}$ $\lambda6191$ band strength.  Dashed
line indicates dwarfs, dotted line indicates giants, while the solid line
indicates those objects with uncertain/unknown luminosities.  Assuming
a significant fraction of the uncertain/unknown objects are giants, the
strength of $\lambda6191$ is not a good indicator of luminosity.
\label{fig8}}
\end{figure}

\begin{figure}
\epsscale{0.8}
\plotone{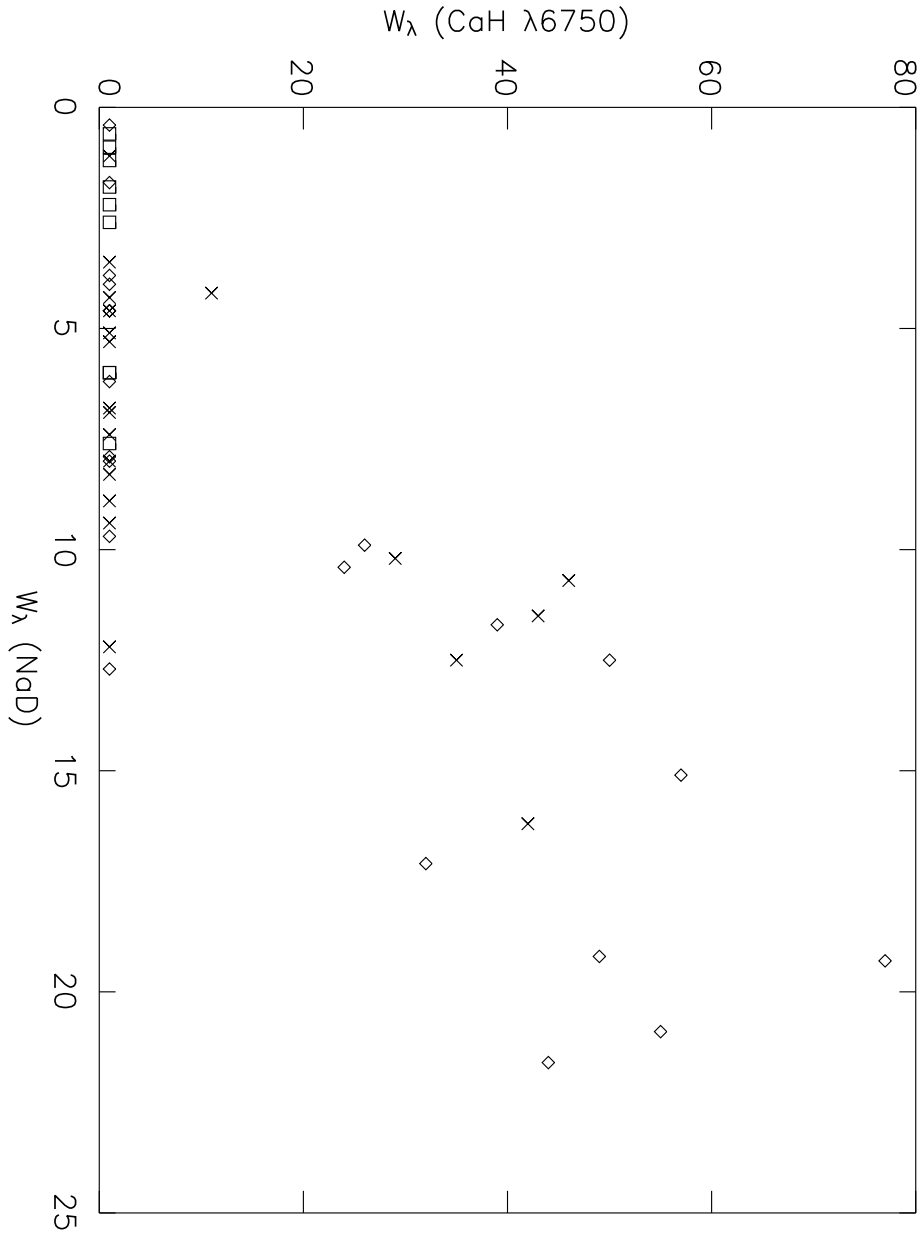}
\caption[fig9.ps]{Plot of W$_{\lambda}$(NaD) vs. W$_{\lambda}$(CaH 
$\lambda6750$).  Same symbols as in Figure~\ref{fig5}.  The objects along
the bottom axis are those with no detectable CaH.  While no confirmed
giants have detected CaH, the large number of uncertain/unknown objects makes
this proposed discriminant uncertain.
\label{fig9}}
\end{figure}

\begin{figure}
\epsscale{0.8}
\plotone{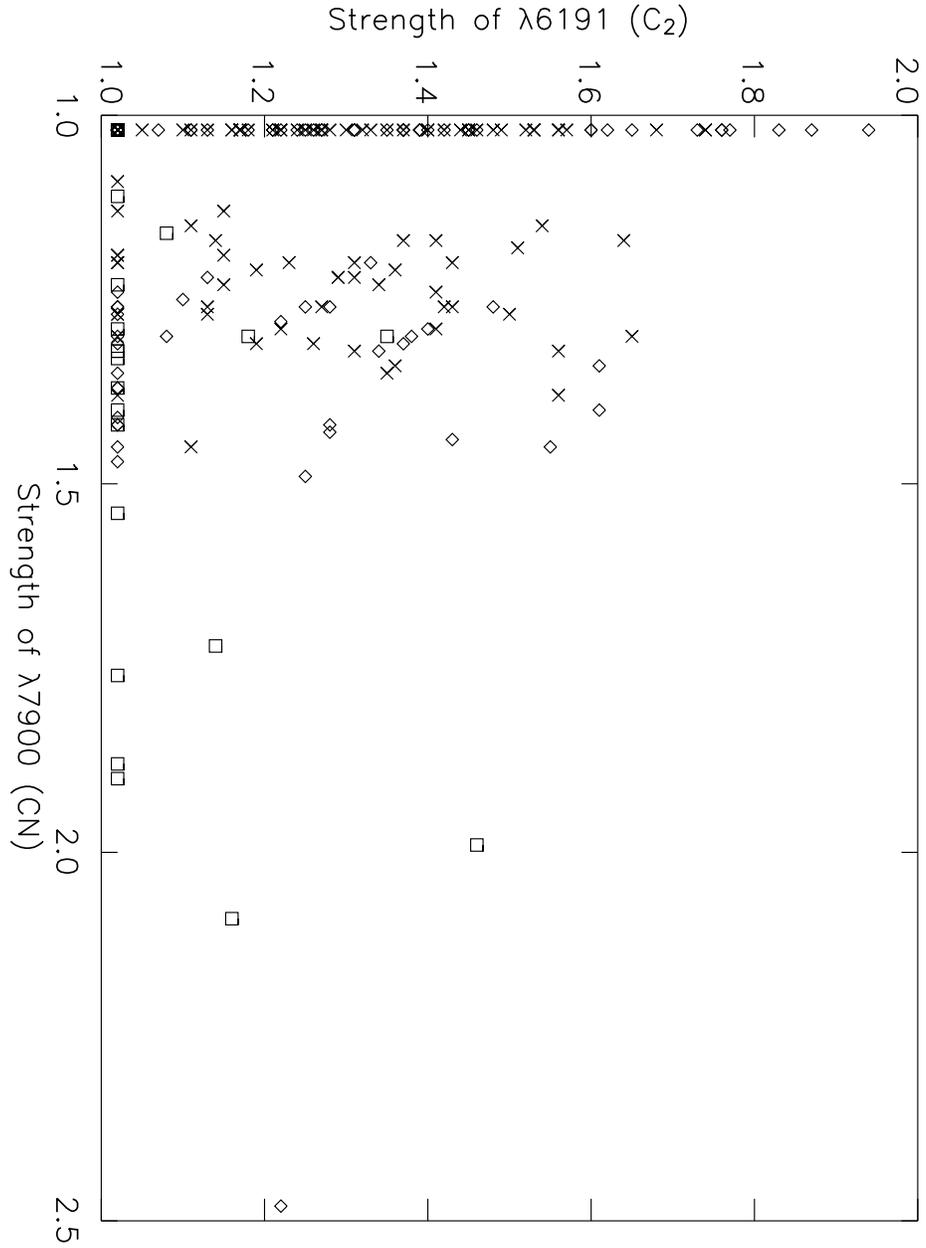}
\caption[fig10.ps]{Plot of the strength of $\lambda7900$ (CN) vs. 
$\lambda6191$ (C$_{2}$).  Same symbols as in Figure~\ref{fig5}.  The objects 
along the axes are those with no detection of CN (y-axis) or C$_{2}$ (x-axis).
The number of confirmed giants is too small to confirm the \citet{gma92}
criterion.  However, the lack of dwarfs with CN strengths greater than 1.5 is
discussed in the text.
\label{fig10}}
\end{figure}

\begin{figure}
\plotone{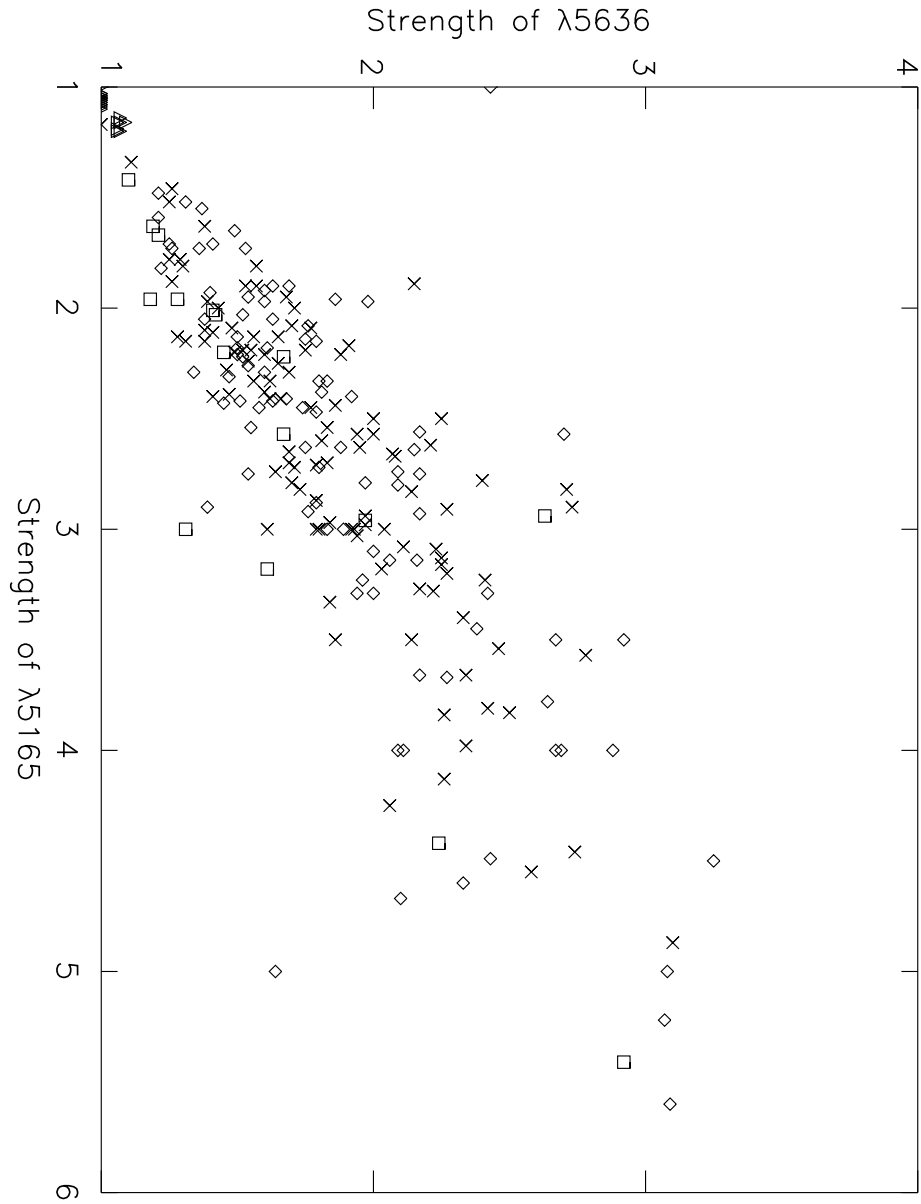}
\caption[fig11.ps]{Plot of the strength of $\lambda5165$ vs. 
$\lambda5636$. Same symbols as in Figure~\ref{fig5}. Note that the 
``F/G Carbon Stars'' fit nicely at the head of the correlation.  There is 
no clear separate of dwarfs and giants.
\label{fig11}}
\end{figure}

\begin{figure}
\epsscale{0.8}
\plotone{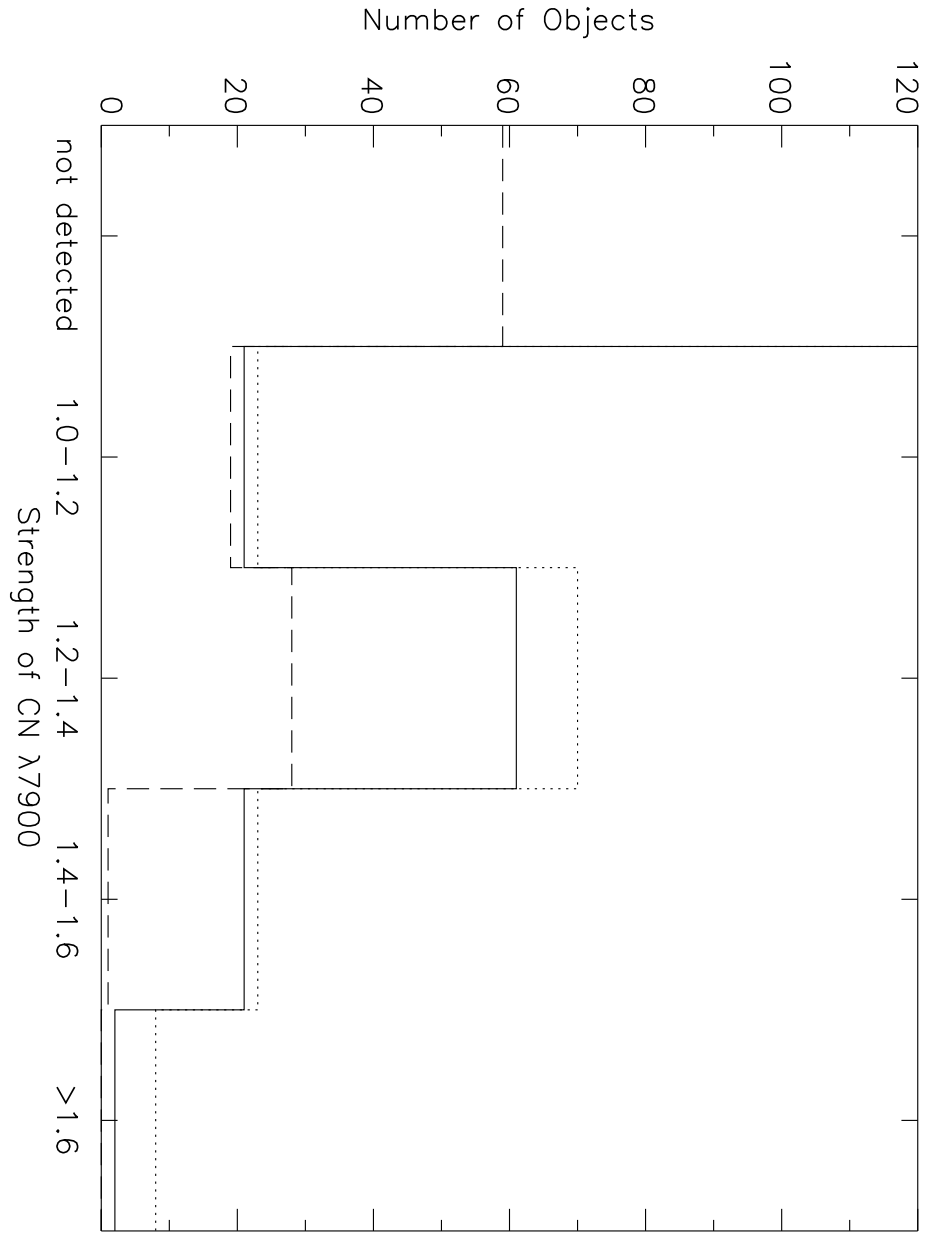}
\caption[fig12.ps]{Histogram of CN $\lambda7900$ strength.  Dashed
line indicates dwarfs, dotted line indicates giants, while the solid
line indicates those objects with uncertain/unknown luminosities.
Those objects with the strongest CN are giants or have uncertain/unknown
luminosities.
\label{fig12}}
\end{figure}


\clearpage
\begin{thebibliography}{}

%\aap = A&A
%\aaps = A&A Supp
%\mnras = MNRAS
%\pasp = PASP
%\apj = ApJ
%\apjs = ApJ Supp
%\aj = AJ

\bibitem[Abazajian et~al.~(2003)]{aba03} Abazajian, K., Adelman, J., Agueros, M. et~al.~2003, 
         \aj, 126, 2081
\bibitem[Alksnis et~al.~(2001)]{alk01} Alksnis, A., Balklavs, A.,
         Dzervitis, U., Eglitis, I., Paupers, O., \&  Pundure, I. 2001,
         Baltic Astron., 10, 1
\bibitem[Bothun et~al.~(1991)]{bem91} Bothun, G., Elias, J. H., MacAlpine, G., 
         Matthews, K., Mould, J. R., Neugebauer, G., \& Reid, I. N. 1991, \aj, 
         101, 2220
\bibitem[Christlieb et~al.~(2001)]{cgw01} Christlieb, N., Green, P. J., 
         Wisotzki, L., \& Reimers, D. 2001, \aap, 375, 366
\bibitem[Cohen (1979)]{coh79} Cohen, M. 1979, \mnras, 186, 837
\bibitem[Dun\'er (1884)]{dun84} Dun\'er, N.-C. 1884, Sur les \'etoiles 
         \`a ~spectres de la troisi\`eme classe (Stockholm: Norstedt)
\bibitem[Fukugita et~al.~(1996)]{fig96} Fukugita, M., Ichikawa, T., Gunn, J. E.,
         Doi, M., Shimasaku, K., \& Schneider, D. P. 1996, \aj, 111, 1748
\bibitem[Grebel (2004)]{gre04} Grebel, E. 2004, private communication
\bibitem[Green et~al.~(1992)]{gma92} Green, P. J., Margon, B., Anderson, S. F., 
         \& MacConnell, D. J. 1992, \apj, 400, 659
\bibitem[Gunn et~al.~(1998)]{gunn98} Gunn, J. E., Carr, M. A., Rockosi, C. M., 
         Sekiguchi, M. et~al.~1998, \aj, 116, 3040
\bibitem[Heber et~al.~(1993)]{hbj93} Heber, U., Bade, N., Jordan, S., \& 
         Voges, W. 1993, \aap, 267, L31
\bibitem[Hendon \& Stone (1998)]{hs98} Hendon, A. A. \& Stone, R. C. 1998, 
         \aj, 115, 296
\bibitem[Hogg et~al.~(2001)]{hog01} Hogg, D. W., Schlegel, D. J., 
         Finkbeiner, D. P., \& Gunn, J. E. 2001, \aj, 122, 2129
\bibitem[J\o rgensen et~al.~(1998)]{jbh98} J\o rgensen, U. G., Borysow, A., \& 
         Hofner, S. 1998, in ASP Conf. Ser. 138, 1997 Pacific Rim Conference 
         on Stellar Atmospheres, ed. K. L. Chan, K. S. Cheng, \& H. P. Singh 
         (San Francisco : ASP), 157
\bibitem[Joyce (1998)]{joyce98} Joyce, R. R. 1998, \aj, 115, 2059
\bibitem[Kilkpatrick et~al.~(1991)]{khm91} Kirkpatrick, J. D., 
         Henry, T. J., \& McCarthy, D. W. 1991, \apjs, 77, 417
\bibitem[Knapp et~al.~(2003)]{knapp03} Knapp, G. R. et~al.~2003, in preparation
\bibitem[Krisciunas, Margon, \& Szkody (1998)]{kms98} Krisciunas, K., Margon, B., \&
         Szkody, P. 1998, \pasp, 110, 1342
\bibitem[Liebert et~al.~(1994)]{lsl94} Liebert, J., Schmidt, G. D., Lesser, M., 
         Stepanian, J. A., Lipovetsky, V. A., Chaffee, F. H., Foltz, C. B., \& 
         Bergeron, P. 1994, \apj, 421, 733
\bibitem[Lowrance et~al.~(2003)]{lkr03} Lowrance, P. J., Kirkpatrick, J. D., 
         Reid, I. N., Cruz, K. L., \& Liebert, J. 2003, \apj, 584, L95
\bibitem[Lupton et~al.~(2001)]{lgi01} Lupton, R., Gunn, J. E., Ivezic, Z., 
         Gnapp, J. R., Kent, S., \& Yasuda, N. 2001, in ASP Conf. Ser. 238, 
         Astronomical Data Analysis Software and Systems X, ed. 
         F. R. Harnden, Jr., A. Primini, \& H. E. Payne (San Francisco : ASP), 269
\bibitem[MacConnell (2003)]{mac03} MacConnell, D. J. 2003, \pasp, 115, 351
\bibitem[Margon et~al.~(2002)]{mar02} Margon, B. et~al.~2002, \aj, 124, 
         1651 (Paper I)
\bibitem[Mould et~al.~(1985)]{msg85} Mould, J. R., Schneider, D. P., Gordon, G. A., 
         Aaronson, M., \& Liebert, J. W. 1985, \pasp, 97, 130
\bibitem[Pier et~al.~(2002)]{pmh02} Pier, J. R., Munn, J. A., Hindsley, R. B., 
         Hennessy, G. S., Kent, S. M., Lupton, R. H., \& Ivezic, Z. 2002, 
         \aj, 125, 1559
\bibitem[Schroeder et~al.~(2002)]{sch02} Schroeder, J. et~al.~2002, \baas, 34, 1126
\bibitem[Smith et~al.~(2002)]{smi02} Smith, J. A., Tucker, D. L., Kent, S. M., et~al.~2002,
         \aj, 123, 2121
\bibitem[Stoughton et~al.~(2002)]{sto02} Stoughton, C. et~al.~2002, \aj, 123, 485 (Early Data Release)
\bibitem[Totten \& Irwin (1998)]{ti98} Totten, E. J. \& Irwin, M. J. 1998, 
         \mnras, 294, 1
\bibitem[Totten et~al.~(2000)]{tiw00} Totten, E. J., Irwin, M. J., \& 
         Whitelock, P. A. 2000, \mnras, 314, 630
\bibitem[Wing \& J\o rgensen (1996)]{wj96} Wing, R. F. \& J\o rgensen, U. G. 
         1996, \baas, 28, 1382 
\bibitem[Yamashita (1967)]{yam67} Yamashita, Y. 1967, Pub. Dom. Astrophys. 
         Obs., XIII, 67
\bibitem[York et~al.~(2000)]{york00} York, D. G., Adelman, J., 
         Anderson, J. E., et~al.~2000, \aj, 120, 1579

\end{thebibliography}
\end{document}